\documentclass[%
 reprint,
 amsmath,amssymb,
 aps,
superscriptaddress]{revtex4-1}

\usepackage{graphicx}
\usepackage{dcolumn}
\usepackage{bm}
\usepackage[compat=1.1.0]{tikz-feynman}
\usepackage{makecell}

\tikzfeynmanset{
empty square/.style={
 /tikzfeynman/square dot,
     /tikz/fill=none
  },
}

\newcommand\T{\rule{0pt}{2.6ex}}       
\newcommand\B{\rule[-1.2ex]{0pt}{0pt}} 

\begin{document}

\preprint{APS/123-QED}

\title{Metastability in networks of nonlinear stochastic integrate-and-fire neurons}

\author{Siddharth Paliwal}
\affiliation{Graduate Program in Neuroscience, Stony Brook University, Stony Brook, NY, 11794, USA}
\author{Gabriel Koch Ocker}
\thanks{co-senior authors}
\email{gkocker@bu.edu}
\affiliation{Department of Mathematics and Statistics, Boston University, Boston, MA, 02215, USA}
\author{Braden A. W. Brinkman}
\thanks{co-senior authors}
\email{braden.brinkman@stonybrook.edu}
\affiliation{Department of Neurobiology and Behavior, Stony Brook University, Stony Brook, NY, 11794, USA}

\date{\today}

\begin{abstract}
Neurons in the brain continuously process the barrage of sensory inputs they receive from the environment. 
A wide array of experimental work has shown that the collective activity of neural populations encodes and processes this constant bombardment of information. 
How these collective patterns of activity depend on single-neuron properties is often unclear. 
Single-neuron recordings have shown that individual neurons' responses to inputs are nonlinear, which prevents a straightforward extrapolation from single neuron features to emergent collective states. 
Here, we use a field-theoretic formulation of a stochastic leaky integrate-and-fire model to study the impact of single-neuron nonlinearities on macroscopic network activity. In this model, a neuron integrates spiking output from other neurons in its membrane voltage and emits spikes stochastically with an intensity depending on the membrane voltage, after which the voltage resets.
We show that the interplay between nonlinear spike intensity functions and membrane potential resets can i) give rise to metastable active firing rate states in recurrent networks, and ii) can enhance or suppress mean firing rates and membrane potentials in the same or paradoxically opposite directions. 

\end{abstract}

\maketitle


\section{\label{sec:Introduction} Introduction}

Populations of neurons can showcase a wide array of complex collective activity patterns that underlie sensory and information processing in the brain. 
Understanding how these macroscopic patterns of activity emerge from the properties of individual neurons has been a central question in neuroscience. 
Recently, metastable activity{---}sharp stochastic or input-driven changes in the firing patterns of the network{---}has come under investigation as a potential means by which populations could flexibly encode sensory information \cite{braden2022MSreview, rossi2023unifiedMS}. 
Metastable states are frequently observed as sub-populations of neurons switching between up (high activity) and down (low activity) states \cite{kajikawa2022udhippocampus, anderson2000udcatvisual, ilan1999udcatvisual, steriade2001ud, bilal2007udvisual}, or transitions between multiple clusters of neurons \cite{schaub2015SSA, miller2010MStaste, recanatesi2022MS}. 
Models of up-down transitions commonly describe the activity in these states as bistable attractors in dynamical models of network activity \cite{tartaglia2017updown, mejias2010udnoise, mattia2012udcortex, jercog2017udtransition, kraynyukova2018ssnbistable, holcman2006emergenceUD}. 
Similarly, models with multiple clusters can produce multistable attractors that describe metastability between multiple states \cite{litwinkumar2012MS, Mazzucato2015MS, mazzucato2019MS, lacamera2019MS, bresloff2010MS}.

Classically, theories for describing emergent macroscopic population activity (e.g., the Wilson-Cowan \cite{wc1972, wc1973} or Amari-Grossberg equations \cite{cohen1983, amari1977, Potthast2013}) are commonly understood as coarse-grained models of the underlying populations of neurons. 
These equations are derived under the implicit assumption of a separation of timescales between the synaptic and intrinsic dynamics of neurons \cite{pinto1996poprate}. 
Additionally, these models describe the mean population activity, ignoring the role that stochastic fluctuations play in shaping the emergent activity of a population of neurons. 
Some models introduce stochasticity into firing rate models by introducing an external fluctuating current input, with variance sometimes self-consistently matched to the firing rates of the network to mimic a recurrent Poisson spiking input. 
However, this phenomenological addition of noise assumes that correlations in spiking activity are negligible, precluding strong coordination within the network.

To properly understand the influence of recurrent spiking activity, we follow recent modeling efforts that employ stochastic field theory representations of the spiking network dynamics  \cite{ocker2017nonlinearHawkes,brinkman2018expGLM,kordovan2020spiketraincumulants,liang2024statistically,brinkman2023non}. 
In particular, we use a stochastic spiking network that incorporates a hard reset of the neurons' membrane potentials after they emit a spike \cite{ocker2023slif}. This hard reset aids in stabilizing network activity, allowing the stochastic neurons to fire guaranteed spikes if their input is large enough yet retain a stable mean firing rate because their membrane potentials are always reset after a spike.

The stability of a network and the possible patterns of collective activity it admits is also intrinsically tied to nonlinearities in neural activity, even at the single-neuron level. 
For example, neurons are often characterized by an intensity (or transfer) function that maps their membrane voltage to an instantaneous firing rate \cite{brunel2014singleneuron, vanmeegen2021microTsSNN, ocker2017nonlinearHawkes, ekelmans2023nonlinearrate, miller2002powerlaw, persi2011powerlawvisual, nicholas2008powerlawvisual, hansel2002catvisual}. 
This intensity function cannot be measured directly; instead, the firing rate as a function of mean membrane potential has been estimated by fitting single-neuron data. The measured nonlinearities are assumed to be a noisy version of the underlying intensity function, giving insight into the appropriate families of intensity functions to use in theoretical and computational models. 
For example, in the primary visual cortex the nonlinearities are fit well by threshold-power law functions of the form $\lfloor V - \theta \rfloor_+^\alpha$, where $\lfloor x \rfloor_+ = x$ if $x > 0$ and $0$ otherwise, and $\alpha$ is an exponent \cite{miller2002powerlaw, persi2011powerlawvisual, nicholas2008powerlawvisual, hansel2002catvisual}. 
Observed exponents $\alpha$ typically fall in the range of 2 $\sim$ 5, larger than the rectified linear units often used to model neural firing in mathematically tractable models \cite{mastrogiuseppe2017intrinsically, kadmon2015transitionchaos, ocker2023slif}. 

\begin{figure}[ht!]
  \begin{center}
  \includegraphics[width=\columnwidth]{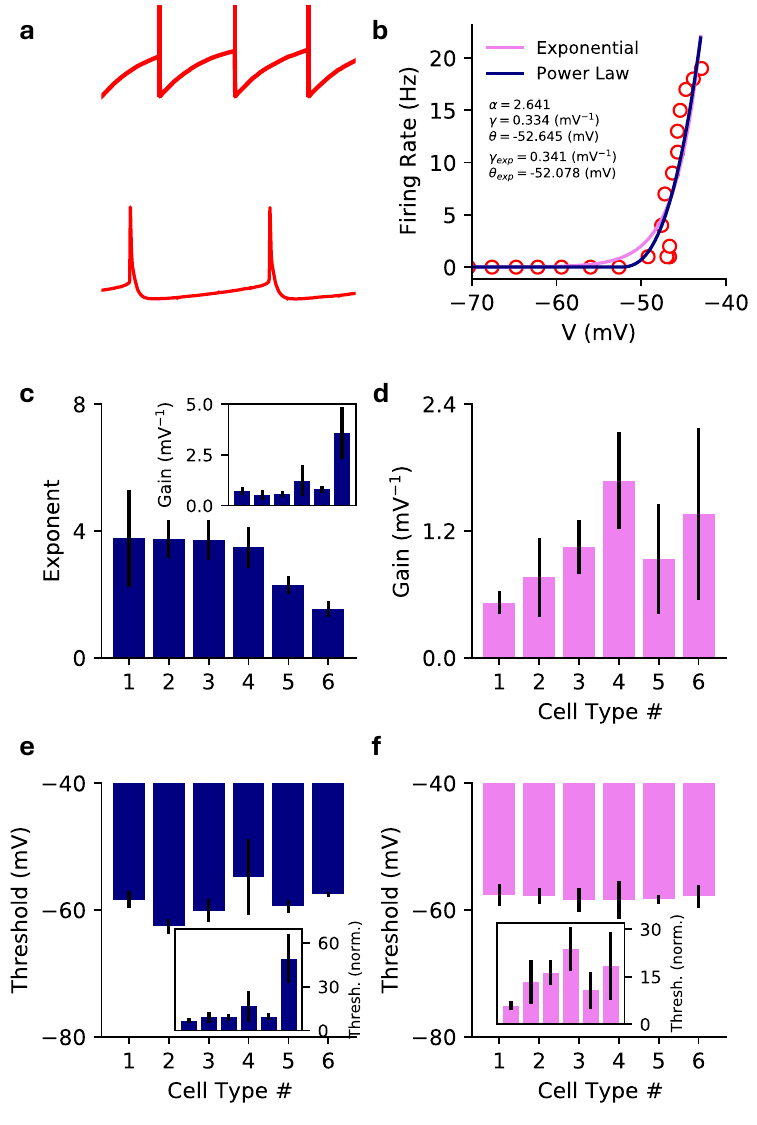}
  \caption{\textbf{a)} (\textit{Top}) Voltage trace showcasing multiple hard-resets after a spike is emitted by a neuron in the model. (\textit{Bottom}) Voltage trace of an example neuron from the cell electrophysiological recording showcasing the full spike generation and post spike hyper-polarization. The model differs from the true dynamics of the cell since the details of the spike generation and hyper-polarization are replaced by the hard-reset. \textbf{b)} Mean membrane potential vs firing rate data for an example cell (empty red circles), with the power law (pink) and exponential (dark blue) fits. The text shows the estimated parameters for both fits. \textbf{c)} The average estimated exponent of the threshold power law fit for different cell types (as identified by Cre line labeling). The inset shows the estimated gain for the fit. \textbf{d)} Same as \textbf{c} but for the exponential fit showing the gain. Cell type 4 is Vip+ neurons, a type of inhibitory cell; all other cell types here are excitatory neurons (Table~\ref{table0}). \textbf{e)} Same as \textbf{c} but showing the threshold and the normalized threshold (inset) for the power law fit. \textbf{f)} Same as \textbf{c} but showing the threshold and the normalized threshold (inset) for the exponential fit. See the main text for the definition of the normalized threshold.} \label{Figure1}
  \end{center}
\end{figure}

The rectifying nature of the threshold power-law stands in contrast with the exponential nonlinearity, an alternative modeling choice that has been used extensively in fitting point process generalized linear models (GLMs) and generalized integrate-and-fire models \cite{simoncelli2004expGLM, paninski2004MLEexpGLM, pillow2005expGLM,badel2007exp, pillow2018expGLM, mensi2011expGLMadap, pozzorini2015expGLM, teeter2018expGLM, brinkman2018expGLM, liang2024statistically}. 
This motivates us to also investigate whether the collective activity generated by such networks approximates the activity of threshold-power law networks.
As shown in Fig.~\ref{Figure1}, we find that both power law and exponential nonlinearities can fit single-neuron intensity functions well across cell types and layers of visual cortex in mouse data from the Allen Institute for Brain Science \cite{teeter2018expGLM}, but we will show that the two classes of nonlinearities exhibit several key differences in the collective activity they can generate.

To this end, we use a stochastic field theory formalization \cite{MSR1_1973, MSR2_1973, MSR3_1976, buiceandchow2015pathintegralsdes, hertz2017pathintegrals, helias2020statfieldtheory, ocker2023slif} of a biologically realistic model of stochastic leaky and fire (sLIF) neurons with nonlinear intensity functions. 
We study its dynamics to compare the impact of threshold power-law and exponential intensity functions on population activity.
We first calculate the mean-field phase diagram of the stochastic spiking network as a function of its synaptic connectivity strength and external input. 
We then show that superlinear threshold-power law intensity functions, coupled with a hard reset of the membrane potential after spiking, enable metastability between i) a quiescent and active state for subthreshold currents and ii) two active states for super-threshold currents. 
In the latter case the networks can stochastically switch between these two active states. 
Moreover, we show that the quiescent steady state is a result of the rectifying nature of the intensity. 
Networks with exponential intensity functions exhibit a monostable regime of continuously varying firing rate or a metastable regime between low and high firing rates.

To capture the impact of fluctuations on population activity, we compute fluctuation corrections to the mean field theory.
These corrections come from the nonlinear firing intensity and the spike reset. 
While the spike reset always suppresses activity, the nonlinear firing intensity promotes activity when the nonlinearity is superlinear and suppresses activity when it is sublinear~\cite{ocker2017nonlinearHawkes}. We study these two effects to determine when fluctuations promote versus suppress activity. 

The paper is organized as follows: in Section \ref{NetworkModel} we introduce the model and the corresponding mean-field equations, with which we obtain the phase diagram for the network. 
In Section \ref{HomogeneousMF}, \ref{EI_MF} we obtain the mean-field phase diagrams for the two nonlinear intensity functions and find that they can have at most two stable steady-state solutions. 
Finally, we analyze the impact of fluctuations on the mean firing rate and membrane potential of the network. We elucidate the impact of two sources of fluctuation correction to the mean-field theory: the spike reset and nonlinear firing intensity, and show that the mean voltage and rate can be promoted, suppressed, or paradoxically corrected in opposite directions, depending on the single-neuron nonlinearity and the network state.

\section{Results} \label{sec:results}
\subsection{Network model and mean-field equations} 
\label{NetworkModel}
Our model comprises a network of $N$ stochastic leaky-integrate and fire neurons (sLIF). 
The membrane dynamics are governed by the stochastic differential equation: 
\begin{align}
    \frac{dV_{i}(t)}{dt} &= - \frac{V_{i}(t)}{\tau} + \sum_{j=1}^{N} J_{ij} \Dot{n}_{j}(t) + \mathcal{E}_{i} - \Dot{n}_{i}(t^{+}) (V_{i}(t)-R) \nonumber \\
    \Dot{n}_{i}(t) dt &\sim {\rm Poiss}\left[\phi(V_{i}(t))dt\right]. \label{SLIF}
\end{align}
where $V_{i}(t)$ is the membrane potential of  neuron $i$ at time $t$, $J_{ij}$ is the connection strength from neuron $j$ to neuron $i$ (equal to $J/(pN)$ with probability $p$, the connection probability, and $0$ otherwise), and $\mathcal{E}_{i}$ is the net external current that each neuron receives. 
The spike trains $\Dot{n}_{i}(t)$ are conditionally Poisson, with instantaneous firing rate $\phi(V_{i}(t))$; i.e., the instantaneous firing rate is a function of the neuron's  membrane potential. 
After the emission of a spike, the last term ($-\Dot{n}_{i}(t^{+}) (V_{i}(t)-R)$) resets the membrane voltage to $R$; see Fig.~\ref{Figure1}a). 
We aim to characterize the steady-states for rectified power-law intensity functions, $\phi(V) = \tau^{-1}\lfloor \gamma(V - \theta) \rfloor^\alpha_{+}$ with exponent $\alpha > 0$ and exponential intensity functions $\phi(V) = \tau^{-1}e^{\gamma(V - \theta)}$, where $\lfloor x \rfloor_+ = x$ if $x > 0$ and $0$ otherwise. 
Here, $\theta$ is the activation threshold in power-law networks and a ``soft'' threshold in exponential networks and $\gamma$ is the gain factor. 
In both cases it is not the threshold at which spikes are guaranteed to occur, but above which the probability of spiking increases sharply.

To motivate the use of these particular nonlinear intensity functions, we fit the average membrane potential versus firing rate to individual cells from the openly available electrophysiology data from the Allen Institute for Brain Science \cite{teeter2018expGLM} (Appendix \ref{appendix:fitting}). 
The dynamics of individual cells differs from the model dynamics where the details of the spiking and hyper-polarization are replaced by the hard-reset (Fig.~\ref{Figure1}b). 
The fitted exponents fall in the range of 2 $\sim$ 4, similar to previous observations~\cite{miller2002powerlaw, persi2011powerlawvisual, nicholas2008powerlawvisual, hansel2002catvisual}, with a mean exponent of 3.2 (Fig. \ref{Figure1}c-e).
As a result, for the threshold-power-law intensity functions we will focus on superlinear exponents.

For the mathematical analyses detailed below it is convenient to non-dimensionalize the model to reduce the parameter set to a few key combinations.
We set $V \rightarrow \gamma^{-1}\hat{V}+R$, $\dot{n} \rightarrow \tau^{-1} \hat{\dot{n}}$, $\phi \rightarrow \tau^{-1} \hat{\phi}$, $\mathcal E \rightarrow \tau^{-1}(\gamma^{-1}\hat{\mathcal E}+R)$, $J_{ij} \rightarrow \gamma  \hat{J}_{ij}$, and $t \rightarrow \hat{t} \tau$, where hats denote dimensionless quantities.
Equivalently, we set $R = 0$, $\tau = 1$, and $\gamma = 1$.
This leaves the dimensionless parameters $\hat{J}$, $\hat{\mathcal E}$ and $\hat{\theta} = \gamma(\theta-R)$.
The last of these parameters can be interpreted as a dimensionless soft threshold.
Its value differs depending on whether we are fitting power-law nonlinearities or exponential nonlinearities to data.
We find in our power-law fits to real data that $\hat{\theta}$ is approximately in the range $1-15$ for most of the cell types we investigated (Fig.~\ref{Figure1}e, inset).
For simplicity, here we set this effective threshold to $\hat{\theta} = 1$ going forward.

In the networks with exponential nonlinearity, Fig.~\ref{Figure1}f shows that $\hat{\theta}$ is in the range $[0,30]$, depending on cell type, with all but one approximately in $[1,15]$. 
Because the exponential nonlinearity has no exponent $\alpha$, we investigate how firing rates change as a function of $\hat{\theta}$.
To lighten notation in the rest of the paper, we omit the hat notation for the non-dimensionalized parameters.

The stochastic dynamics in \eqref{SLIF} gives rise to a probability distribution for the membrane potential and firing rate for the network. 
The joint moment generating functional (MGF) $Z[\Tilde{j}, j, \Tilde{h}, h]$ for membrane potentials and spike trains can be expressed as a path integral using the Martin-Siggia-Rose-De Dominicis-Janssen (MSRDJ) formalism:
\begin{widetext}   
\begin{align}
    Z[\Tilde{j}, j, \Tilde{h}, h] &= \int \mathcal{D}\tilde{V} \, \mathcal{D} V \mathcal{D} \tilde{n} \, \mathcal{D} \Dot{n} \, \exp \left(-S[\Tilde{V},V,\Tilde{n},\Dot{n}] + \Tilde{j} \cdot V + j \cdot \Tilde{V} + \Tilde{h} \cdot \Dot{n} + h \cdot \Tilde{n} \right), \nonumber \\
    S\left[\tilde{V},V,\tilde{n},\dot{n}\right] &= \sum_{i=1}^{N} \int_{-\infty}^{\infty} dt~ \Biggl\{ \tilde{V_{i}}(t) \left[ {\Dot{V}_{i}(t)} + V_{i}(t) - \mathcal E_{i} + \Dot{n}_{i}(t^{+}) V_{i}(t) - \sum_{j=1}^{N} J_{ij} \dot{n}_{j}(t) \right] \nonumber \\
    & \hspace{6.0cm} + \tilde{n}_{i}(t) \dot{n}_{i}(t) -\left(e^{\tilde{n}_{i}(t)} -1 \right) \phi(V_{i}(t)) \Biggr\}. \label{MGF_Full}
\end{align}
\end{widetext}
where $\Tilde{x} \cdot y = \sum_{i} \int  dt~\Tilde{x}_{i}(t) y_{i}(t)$, $\Tilde{V}, \Tilde{n}$ are the response variables, which can be interpreted as noise variables driving the membrane potential $V$ and firing rate $\Dot{n}$, and $\{\Tilde{j}, j, \Tilde{h}, h\}$ are source fields. 
Moments of the joint probability distribution can be obtained by taking derivatives of the MGF with respect to the source fields.
Generating functionals are a powerful tool for studying stochastic dynamics \cite{buiceandchow2015pathintegralsdes, buice2013fieldtheoryreview, hertz2017pathintegrals, helias2020statfieldtheory}. 
In neuroscience,
these functional methods have previously been applied to firing rate networks, e.g., \cite{sompolinsky1988chaosRNN, kadmon2015transitionchaos, vanMeegen2021ratepathintegral, keup2021transchaoticRNN} and coarse-grained networks of neural activity, e.g., \cite{buice2013finitesizeSNN, bresslofff2014transitionmarkov, bressloff2015pathintegral, buice2010pathintegral, stapmanns2020SCnonlinearpathintegral}.
Recent work has also applied such methods to point process networks with linear voltage dynamics, without a hard reset of the voltage after a spike \cite{ocker2017nonlinearHawkes,brinkman2018expGLM,kordovan2020spiketraincumulants,liang2024statistically}. 
In this work, we follow \cite{ocker2023slif} in applying this formulation to spiking networks with hard resets.

For random synaptic connections $J_{ij}$ we can average the MGF over realizations of the synaptic connections to derive an effective dynamical description of the population dynamics.  
In the weak-coupling regime in which the weights $J_{ij}$ scale as $1/N$, one can ignore the higher cumulants of the connectivity matrix in the large-$N$ limit and describe the collective dynamics of the network using only the mean connection strength $J$. 
If we assume that there are $M$ homogenous populations of neuron, this mean-field treatment reduces the $N${-}dimensional dynamics to an effective $M${-}dimensional description of the population statistics within each cluster corresponding to the action 
\begin{widetext}
\begin{align} \label{MGF_DO}
 S[\tilde{V},V,\tilde{n},\dot{n}] &= \sum_{\mu=1}^{M} \int_{-\infty}^{\infty} dt~ \Biggl\{ \tilde{V}_{\mu}(t) \left[ \Dot{V}_{\mu}(t) + V_{\mu}(t) - \mathcal{E}_{\mu} + \Dot{n}_{\mu}(t^{+}) V_{\mu}(t) - \sum_{\nu=1}^{M} J_{\mu \nu} \langle \Dot{n}_{\nu}(t) \rangle \right] \nonumber \\
    & \hspace{6.0cm} + \tilde{n}_{\mu}(t) \dot{n}_{\mu}(t) -\left(e^{\tilde{n}_{\mu}(t)} - 1 \right) \phi(V_{\mu}(t)) \Biggr\}. 
\end{align}
\end{widetext}
where the Greek indices label the different clusters and $\langle \Dot{n}_{\nu} \rangle$ is the population-averaged activity that needs to be determined self-consistently from the mean-field equations of motion. 
This decouples the neurons in the network, such that the average behavior of an individual neuron in a cluster depends only on the mean synaptic field $\sum_{\nu=1}^{M} J_{\mu \nu} \langle \dot{n}_{\nu} \rangle$. 
This description is exact in the $N \to \infty$ limit \cite{robert2016dynamics}.
The MGF corresponding to \eqref{MGF_DO} can then be used to obtain the deterministic mean-field approximation for the population membrane potential and firing rate:
\begin{align} \label{MFT}
    \frac{d\Bar{V}_{\mu}}{dt} &= -\Bar{V}_{\mu} + \mathcal{E}_{\mu} + \sum_{\nu=1}^{M} J_{\mu \nu} \Bar{n}_{\nu} - \Bar{n}_{\mu} \Bar{V}_{\mu}, \nonumber \\
    \Bar{n}_{\mu} &= \phi(\Bar{V}_{\mu}).
\end{align}
Unlike standard dynamical mean-field treatments of firing rate networks, Eq.~\ref{MFT} is not exact. 
The stochasticity of spiking remains in Eq.~\ref{MGF_DO} and influences the dynamics of the population-averaged membrane potentials $V_\mu(t)$ through the spike reset $\dot{n}_\mu(t^+)V_\mu(t)$. 
Nevertheless, deterministic mean-field approximations often paint a good qualitative picture of the dynamics of the system and we will first investigate the different phases of network activity within this approximation.
Then, in Section \ref{fluctuationImpact}, we account for the influence of spiking fluctuations on network activity by expanding around steady-state solutions of the mean-field equations to calculate Gaussian (perturbative one-loop) corrections.

Alternatively, we can take advantage of the hard reset to estimate the exact firing rate of the networks using renewal theory. 
For self-averaging networks, the density in \eqref{MGF_DO} factorizes over clusters. 
Each neuron ($\mu \in M$) spikes independently given a self-consistent mean-field input. 
When the network reaches a steady-state, the self-consistent input becomes time-independent, in which case the spiking dynamics can be treated as a renewal process due to the hard reset of the membrane potential after spike emission. 
After each spike, the membrane potential asymptotically evolves to the net input from the reset potential (=0 here):
\begin{equation}
    V_{\mu}(t) = C_{\mu} (1 - e^{-t}),
\end{equation}
where $C_{\mu}=\mathcal{E}_{\mu} + \sum_{\nu=1}^{M} J_{\mu \nu} \langle \dot{n}_{\nu} \rangle$ is the net input that the neuron receives with $\langle \Dot{n}_{\nu} \rangle$ to be determined self-consistently. 
For an arbitrary nonlinearity $\phi(V(t))$ the inter-spike interval density $p_{\mu}(s)$ is the product of the instantaneous firing probability at time $s$ and the survival probability until time $s$:
\begin{equation}
    p_{\mu}(s) = \phi(V_{\mu}(s)) \exp{\Big( -\int_{0}^{s} \phi(V_{\mu}(t)) dt \Big)}
\end{equation}
The mean inter-spike interval for each cluster is $\langle s_{\mu} \rangle = \int_{0}^{\infty} ds ~ s \, p_{\mu}(s)$. The self-consistent firing rate is the inverse of the mean inter-spike interval: $\langle \Dot{n}_{\mu} \rangle = 1 / \langle s_{\mu} \rangle$, and the roots of this system of equations allow us to estimate the steady-state firing rates for the network.
\begin{figure*}
  \begin{center}
  \includegraphics[width=\textwidth]{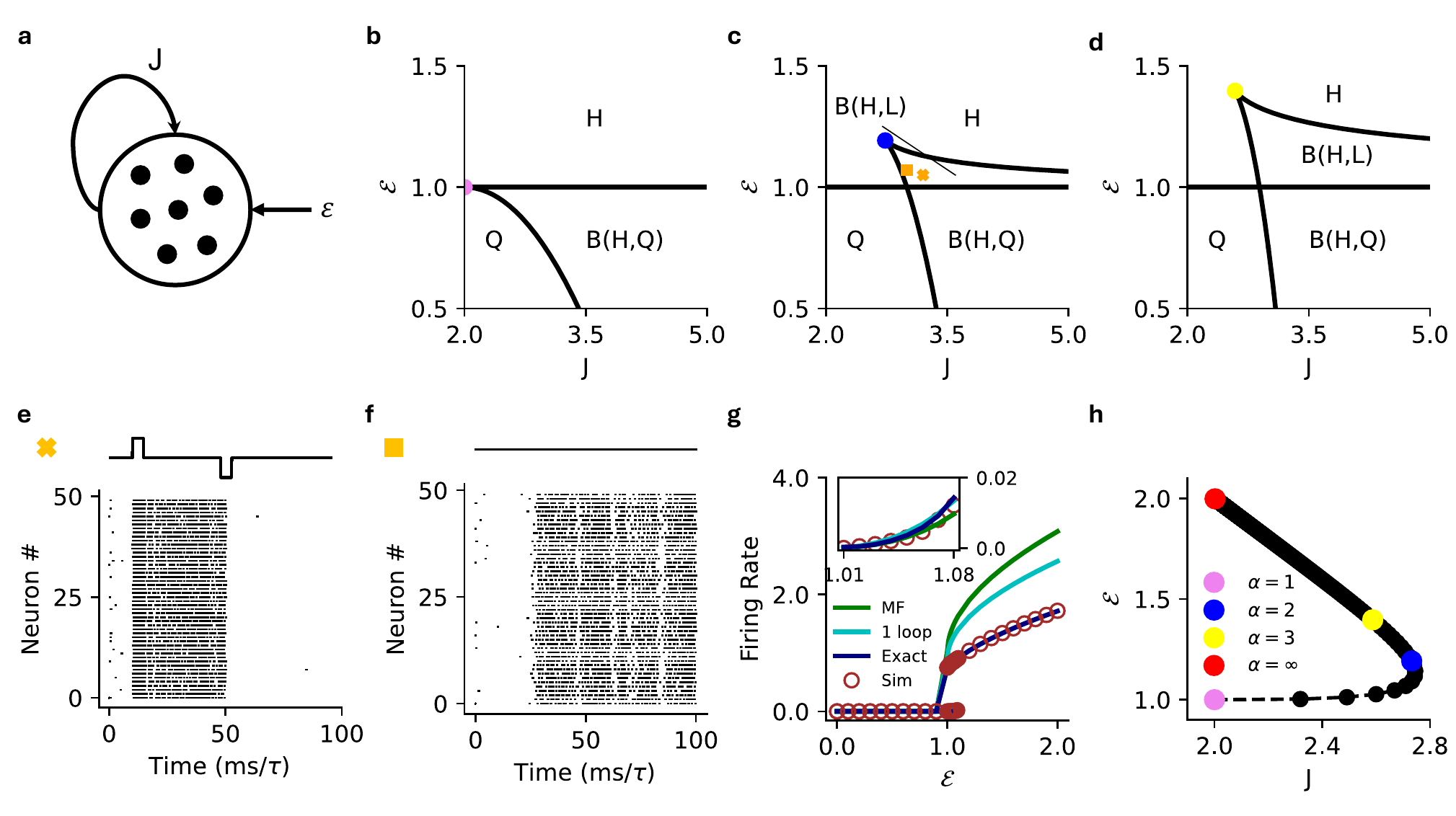}
  \caption{\textbf{a)} Schematic for the homogeneous network. The network has a coupling strength $J$ and receives an external current $\mathcal{E}$. \textbf{b)} Mean-field (MF) phase diagram for the homogeneous network with $\alpha=1$ in the $\mathcal{E}{-}J$ plane, separating the bistable (B{--}H, Q) high firing rate and quiescent region from the monostable high (H) firing rate and quiescent (Q) regions. \textbf{c)} Same as \textbf{b} for $\alpha=2$ separating the bistable (B{--}H, Q) high firing rate and quiescent, bistable (B{--}H, L) high and low firing rate regions from the monostable high (H) firing rate and quiescent (Q) regions. \textbf{d)} Same as \textbf{c} for $\alpha=3$. \textbf{e)} Raster plot for the stochastic spiking network ($\alpha=2$) at the parameter marked with cross in panel \textbf{c} ($J=3.2$, $\mathcal{E}=1.05$), illustrating input driven transition between the two active states. \textbf{f)} Raster plot for the stochastic spiking network ($\alpha=2$) at the parameter marked with square in panel \textbf{c} ($J=3.0$, $\mathcal{E}=1.07$), illustrating stochastic transition between the two active states. \textbf{g)} MF (green), 1-loop (cyan) and renewal theory (dark blue) firing rate predictions compared to simulations (brown) for fixed $J=3.0$ ($\alpha=2$). Inset highlights the low firing rate state in the black square. \textbf{h)} MF phase diagram that depicts that the critical point continuously increases with the exponent of the intensity and reaches a limiting value for arbitrarily large $\alpha$. Violet: $\alpha = 1$, Blue: $\alpha=2$, Yellow: $\alpha=3$, Red: $\alpha \rightarrow \infty$.} \label{Figure2}
  \end{center}
\end{figure*} \label{Figure2}

\subsection{Homogeneous Networks} \label{HomogeneousMF}
We are interested in examining the steady states of the mean-field equations \eqref{MFT}. 
For simplicity, we start with a homogeneous excitatory network of $N$ coupled neurons (Fig. \ref{Figure2}a). 
In this case, the mean-field theory is one-dimensional and a phase diagram can be obtained by investigating the fixed points of the mean-field dynamics. 
The phase diagram will be a function of the external input ($\mathcal{E}$) and the mean coupling strength ($J$).

\subsubsection{Threshold Power-Law Intensity}
We first investigate population activity of threshold power-law networks. The mean-field equation of motion simplifies to 
\begin{equation}
    \frac{d\Bar{V}}{dt} = -\Bar{V} + \mathcal{E} + (J - \Bar{V}) \lfloor\Bar{V} - 1\rfloor_+^{\alpha},
\end{equation}
where we have fixed the threshold at $\theta=1$ and gain at $\gamma = 1$ for simplicity.
The intensity function is zero below threshold so $V=\mathcal{E}$ is the only possible sub-threshold solution, which does not exist if $\mathcal E > \theta$. 
Above threshold, one can show that for any value of the exponent $\alpha$, the above equation can have a maximum of 3 steady-state solutions, giving rise to bistability in the mean-field dynamics (see Appendix \ref{appendix:sspowerlaw}). 
Bistability in the mean-field dynamics corresponds to metastability in the stochastic spiking network dynamics.
It was previously shown in Ref.~ \cite{ocker2023slif} that with a threshold linear intensity function ($\alpha = 1$), the homogeneous network can either be monostable or metastable between an active and quiescent state (Fig. \ref{Figure2}b). 
It is not possible to explicitly calculate the full phase diagram solution for arbitrary $\alpha$, but the cases $\alpha = 2, 3$ are analytically tractable and we give the full solutions here (Fig.~\ref{Figure2}c,   d). 
For $\alpha > 3$ we can still characterize several general properties of the phase diagram.

We begin with the case $\alpha=2$.
The mean-field equation of motion \eqref{MFT} for the threshold-quadratic intensity function is a cubic equation (assuming $\bar{V}$ is above threshold):
\begin{align}
    \Bar{V}^{3} - \Bar{V}^{2} (J+2) + \Bar{V}(2J+2) - (J + \mathcal{E}) &= 0.
    \label{eqn:cubicV}
\end{align}
While one can solve for the roots of this equation, we are more interested in the phase diagram of the network in the $\mathcal{E}{-}J$ plane, which motivates us to find the boundaries between cases for which we have multiple real solutions above threshold. 
We can identify the phase boundaries by analyzing the discriminant, which determines the boundary between a single real solution and three real solutions of Eq.~(\ref{eqn:cubicV}). 
The discriminant is a quadratic in $\mathcal E$ whose roots are functions of $J$, yielding the mean-field boundaries between bistable and monostable phases; see Appendix~\ref{appendix:sspowerlaw} for details.
Analogous to the threshold-linear intensity function, the mean-field dynamics are bistable for sufficiently large values of $J$. 
However, as mentioned above, in threshold-linear networks the bistability is only between a quiescent state and an active state that exists only for subthreshold input $\mathcal E$. 
The superlinear networks retain this regime, but can also exhibit bistability for superthreshold input, and in this region, the bistable state has two active steady-state solutions (Fig. \ref{Figure2}c). This allows for either forced or spontaneous switching between two corresponding metastable active states in the full stochastic network (Fig. \ref{Figure2}e, f).

Turning to $\alpha=3$, the mean-field equation of motion \eqref{MFT} for the threshold-cubic intensity function is a quartic polynomial when $\bar{V}$ is above threshold:
\begin{align}
    \Bar{V}^{4} + b\Bar{V}^{3} - c\Bar{V}^{2} + d \Bar{V} + f &= 0
\end{align}
where the coefficients $\{b,c,d, f\}$ are $b = -(J+3),~c = 3(J+1),~d = -3J,~f = (J + \mathcal{E})$.
As in the previous case, we are interested in identifying phase boundaries, which we can again identify by investigating the discriminant of the quartic equation, this time a cubic in $\mathcal{E}$; we give the details in Appendix~\ref{appendix:sspowerlaw}. 
We again find that only two stable fixed points exist above threshold, meaning the bistable state can have two active steady-state solutions for superthreshold input (Fig. \ref{Figure2}d). 
Furthermore, we observe that increasing the exponent $\alpha$ from $2$ to $3$ expands the region in which both the bistable states are active (Fig. \ref{Figure2}c vs d).

For the homogeneous network, the membrane dynamics can be treated as a renewal process where after each spike the membrane potential evolves as: $V(t) = C(1-e^{-t})$ with $C=\mathcal{E} + J \langle \Dot{n} \rangle$. 
The mean firing rate can then be estimated by numerically finding the roots of (see Appendix \ref{appendix:renewalFR} for details):
\begin{equation}
    \langle \Dot{n} \rangle = \Bigg( \int_{\ln{\frac{C}{C-1}}}^{\infty} ds~ s \big(C(1-e^{-s}) - 1 \big)^{\alpha} e^{-I_{p}(s|C)}\Bigg)^{-1}
\end{equation}
where
\begin{align*}
    I_{p}(s|C) =& \frac{(C(1-e^{-s}) - 1)^{\alpha+1}}{(C-1)(\alpha+1)} \nonumber \\
    &\times \; {}_2F_{1}\Bigg(1, \alpha+1, \alpha+2, \frac{(C(1-e^{-s}) - 1)}{(C-1)}\Bigg)
\end{align*}
and ${}_2F_{1}(a,b,c,x)$ is the hypergeometric function. 
The subscript $p$ refers to the power-law intensity function ($\alpha$ is its exponent). Fig.~\ref{Figure2}g shows the firing rate predictions of the mean-field theory, the one-loop correction (discussed in Section~\ref{fluctuationImpact}), and renewal theory compared to simulations.

While analytic results are difficult to obtain for $\alpha > 3$, we can show that there are at most two stable fixed points above threshold, meaning the mean-field dynamics can be at most bistable (Appendix \ref{appendix:sspowerlaw}). 
We can also determine the coordinates of the `cusp' of the phase diagram where the two phases boundaries meet for arbitrary $\alpha$, finding that it traces a non-monotonic path through the $J{-}\mathcal E$ plane (Fig. \ref{Figure2}h). 
Our result suggests that for $\alpha \gtrsim 2$ an increase in exponent of the intensity function expands the region of bistability where two active metastable steady-states can coexist. 

In all cases the bistable region of the $(J, \mathcal E)$ plane is subdivided in two. Below the threshold $\mathcal E=1$, there exists a quiescent steady state and a high-rate steady state. 
Above threshold, the quiescent state becomes a low-rate state. 
Both active states are metastable in the full stochastic network, which can stochastically transition between them (Fig. \ref{Figure2}f). 
If the states are far apart so that spontaneous transitions are unlikely, a transition can still be induced by stimulation (Fig. \ref{Figure2}e).  

\begin{figure}
  \begin{center}
  \includegraphics[width=\columnwidth]{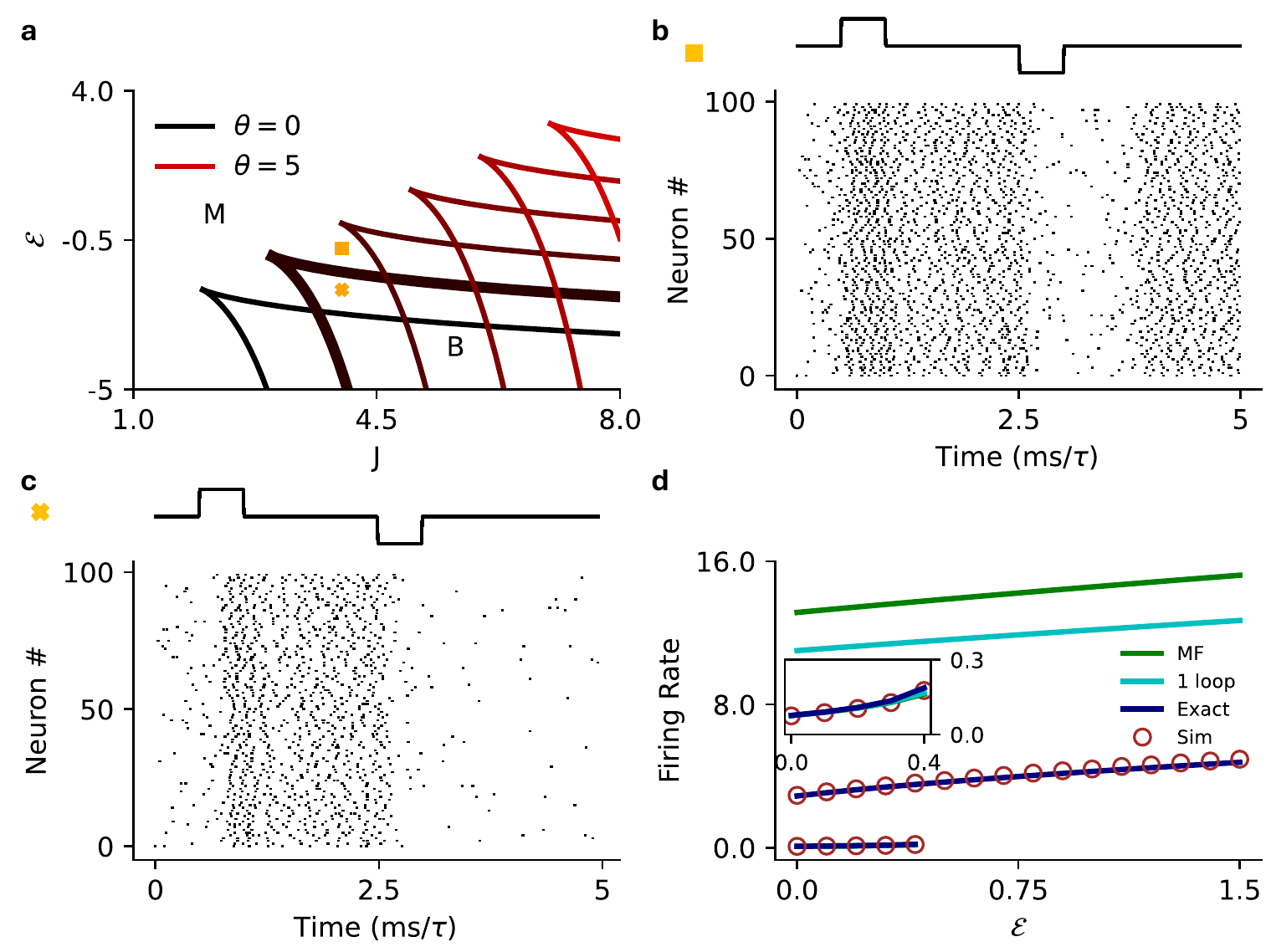}
  \caption{\textbf{a)} MF phase diagram for the exponential intensity function separating the bistable (B) regime from the monostable (M) regime for various values of $\theta \in [0, 5]$ \textbf{b)} Raster plot for the stochastic spiking network ($\theta=1$) at the parameter marked with square in panel \textbf{a} ($J=4.0$, $\mathcal{E}=-0.75$), illustrating the monostable states. \textbf{c)} Raster plot for the stochastic spiking network ($\theta=1$) at the parameter marked with cross in panel \textbf{a} ($J=4.0$, $\mathcal{E}=-2.0$), illustrating input driven transition between the the two active states. \textbf{d)} MF (green), 1-loop (cyan) and renewal theory (dark blue) firing rate predictions compared to simulations (brown) for fixed $J=6.0$ ($\theta=3$). Inset highlights the low firing rate state in the main plot.} \label{Figure3}
  \end{center}
\end{figure} 
\subsubsection{Exponential intensity functions}
For an exponential intensity function, $\phi(V) = e^{V-\theta}$, we find that the mean-field dynamics can be bistable between two states or monostable, similar to the threshold power law networks. 
However, the exponential nonlinearity is never zero for any finite input, so the network does not admit a quiescent state. 
We show in Appendix~\ref{appendix:ssexponential} that the mean-field dynamics of the homogeneous exponential network are bistable if:
\begin{equation}
\begin{aligned}
    J &> \theta + 2, \\
    \mathcal{E} &> J - (1 - W_{0} ) \left[1 + \exp\left(J-1-\theta+W_{0} \right) \right], \\
    \mathcal{E} &< J - (1 - W_{-1}) \left[1 + \exp\left(J-1-\theta+W_{-1}\right) \right],
\end{aligned}
\end{equation}
where $W_{0,-1} \equiv W_{0,-1}(-e^{\theta + 1 - J})$ is the Lambert W function. 
Figure \ref{Figure3}a shows the mean-field bistability boundaries for various values of $\theta \in [0, 5]$. 
We see these states in the stochastic spiking network as well, where the network can either be monostable (Fig. \ref{Figure3}b) or metastable (Fig. \ref{Figure3}c). 

We can also estimate the exact firing rates for the exponential intensity using renewal theory. 
In this case, the self-consistent firing rates are roots of (see Appendix \ref{appendix:renewalFR} for details):
\begin{equation}
    \langle \Dot{n} \rangle = \Bigg( \int_{0}^{\infty} ds \, s \, \exp \left[C (1-e^{-t})- \theta-I_{e}(s|C) \right] \Bigg)^{-1},
\end{equation}
where
$I_{e}(s|C) = \exp\left( C-\theta \right) [-{\rm Ei}(-C e^{-s}) + {\rm Ei}(-C)]$
and 
${\rm Ei}(x) = -\int_{-x}^{\infty} dt~e^{-t} / t$ is the exponential integral. Fig.~\ref{Figure3}d shows the firing rate predictions for a homogeneous network with an exponential intensity function. 

\subsection{Excitatory-Inhibitory Networks} \label{EI_MF}
\begin{figure}
  \begin{center}
  \includegraphics[width=\columnwidth]{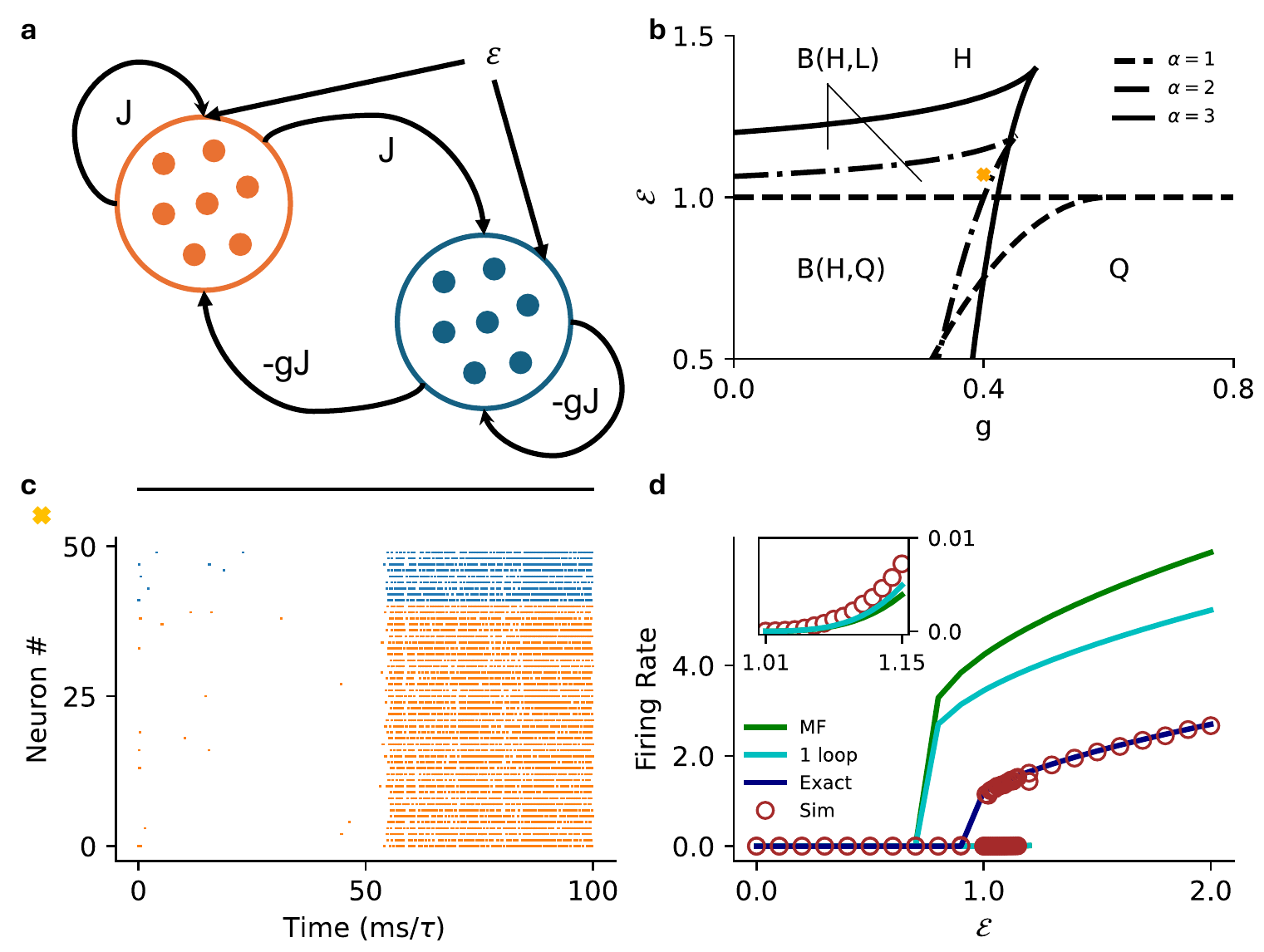}
  \caption{\textbf{a)} Schematic for the excitatory-inhibitory (EI) network (E: Orange, I: Blue). The excitatory cluster has a coupling strength $J$, the relative strength of inhibition is $g$, and both populations receive an external current $\mathcal{E}$. \textbf{b)} MF phase diagram for threshold power law intensity in the $\mathcal{E}{-}g$ plane for $\alpha=1$ (dashed), $\alpha=2$ (dot dashed) and $\alpha=3$ (solid). The phase are same as labeled in Fig. \ref{Figure2}. \textbf{c)} Raster plot for the stochastic spiking network ($\alpha=2$) at $J=5.0$, $g=0.4$, and $\mathcal{E}=1.07$ illustrating stochastic transition between the two active states. \textbf{d)} MF (green), 1-loop (cyan) and renewal theory (dark blue) firing rate predictions compared to simulations (brown) for fixed $J=5.0$, $g=0.4$ ($\alpha=3$). Inset highlights the low firing rate state in the main plot.} \label{Figure4}
  \end{center}
\end{figure}

Neuronal networks in the brain are not homogeneous. 
Realistic networks follow Dale's law: there are both excitatory and inhibitory neurons in these networks. 
To look at more biologically realistic networks, here we consider an excitatory-inhibitory (EI) network with pulse coupling (Fig. \ref{Figure4}a). 
We introduce an additional parameter ($g$) that quantifies the relative strength of inhibition such that the mean connectivity matrix has the following structure:
\begin{equation}
    \begin{bmatrix}
       J_{EE} & J_{EI} \\
       J_{IE} & J_{II} \\   
    \end{bmatrix}
    =
    \begin{bmatrix}
       J & -gJ \\
       J & -gJ \\   
    \end{bmatrix}
\end{equation}
If both populations receive the same input, the mean-field theory for this EI network reduces to the one-dimensional model with the replacement $J \rightarrow J(1-g)$ in the homogeneous network results. 

In Fig.~\ref{Figure4} we show the results for threshold power law intensities in the $\mathcal E{-}g$ plane (Fig. \ref{Figure4}b), along with a raster plot of a simulation to confirm the full stochastic network exhibits metastability (Fig. \ref{Figure4}c). 
We compare the mean-field predictions of the firing rates with the one-loop corrections (discussed in the next section), and the estimates from renewal theory and the simulations (Fig. \ref{Figure4}d).

\begin{figure}
  \begin{center}
  \includegraphics[width=\columnwidth]{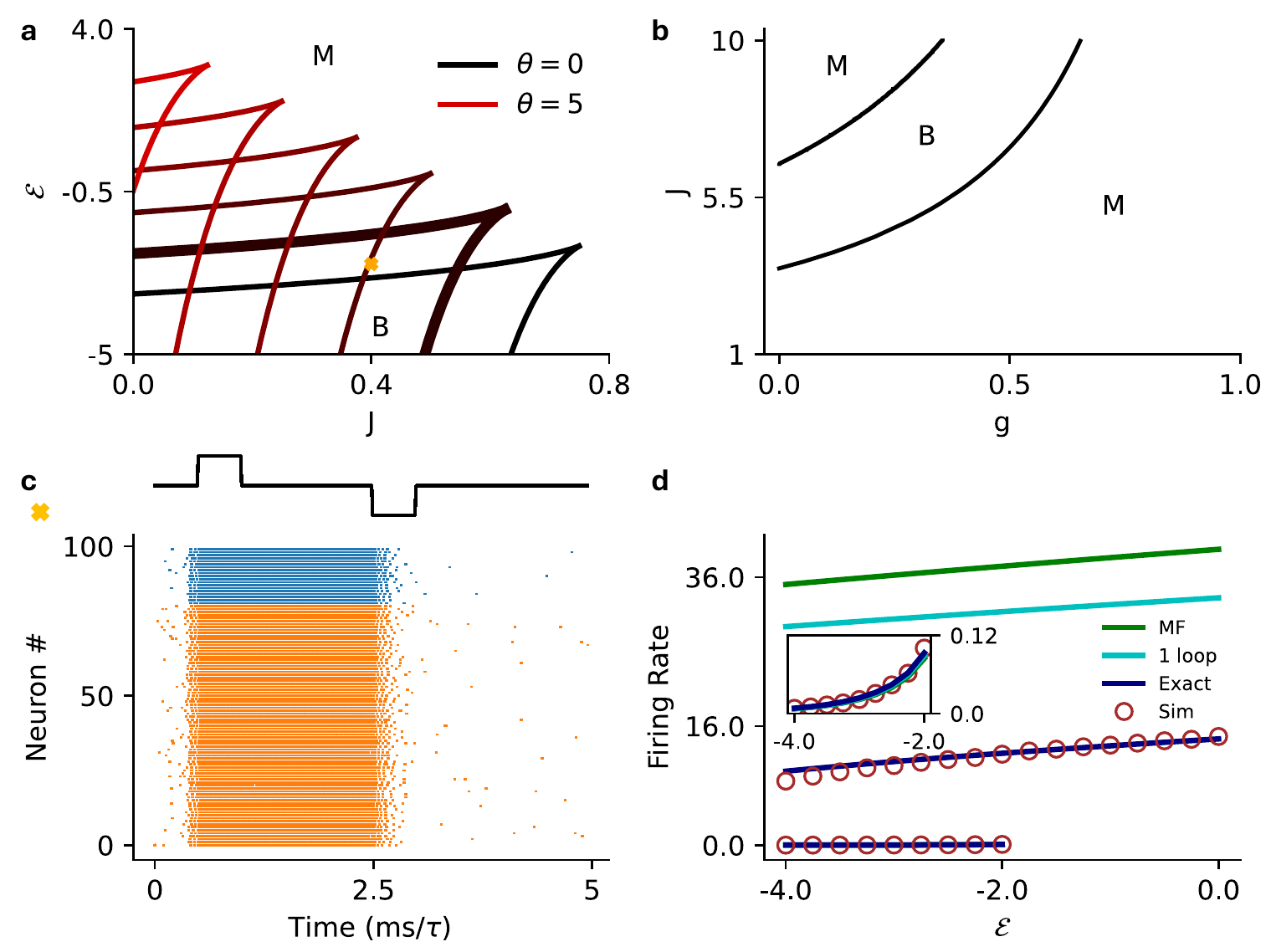}
  \caption{\textbf{a)} MF phase diagram for the exponential intensity in the $\mathcal{E}{-}g$ plane for various values of $\theta \in [0, 5]$. The phase are same as labelled in Fig. \ref{Figure3}. \textbf{b)} Same as \textbf{a} but in the $J{-}g$ plane for $\theta=1$. \textbf{c)} Raster plot for the stochastic spiking network ($\theta=1$) at the parameter marked with cross in panel \textbf{a} ($J=8.0$, $g=0.4$, and $\mathcal{E}=-2.5$), illustrating input driven transition between the the two active states. \textbf{d)} MF (green), 1-loop (cyan) and renewal theory (dark blue) firing rate predictions compared to simulations (brown) for fixed $J=8.0$ and $g=0.4$ ($\theta=1$). The inset shows the low firing rate state in the main figure.} \label{Figure5}
  \end{center}
\end{figure}

In Fig.~\ref{Figure5} we show the corresponding results for E-I networks with exponential nonlinearity, this time in the $J{-}g$ plane (Fig. \ref{Figure5}b), along with demonstration of input-driven metastability (Fig. \ref{Figure5}c) and the comparisons of mean-field, one-loop, renewal theory, and simulation estimates of the firing rates (Fig. \ref{Figure5}d).

\subsection{Impact of fluctuations on the mean activity} 
\label{fluctuationImpact}
While the mean-field theory provides a reasonably accurate qualitative description of the network dynamics and phase transitions in the network, it is quantitatively inaccurate (e.g., Figs.~\ref{Figure2}g, \ref{Figure3}-\ref{Figure5}d) because it ignores all fluctuations. 
These fluctuations impact the mean activity due to i) the nonlinear spike reset and ii) the nonlinear intensity function. 
In isolation, these have opposite effects: the spike reset suppresses activity~\cite{ocker2023slif}, while a concave-up intensity function promotes activity~\cite{ocker2017nonlinearHawkes}. 
We next ask: in the presence of both sources of nonlinearities, do fluctuations enhance firing or suppress it?

To compare the effect of each nonlinearity on the mean voltage vs the mean firing rate, we calculated perturbative one-loop corrections to the mean-field predictions (c.f. the self-consistent one-loop theory of \cite{ocker2023slif}). 
In the path integral formalism these corrections can be evaluated by perturbatively expanding the MGF under a Gaussian approximation. 
This can be achieved by the use of a diagrammatic expansion (\cite{ocker2023slif}; Appendix \ref{appendix:feynmanrules}). 
These Feynman diagrams are constructed from edges and vertices.

The edges in the Feynman diagrams correspond to the linear responses of each variable ($V, \Dot{n}$) to fluctuations in either ($V, \Dot{n}$) -- also called propagators. 
Since we have 2 each of configuration and response variables, they give rise to 4 propagators in the model, given here in the frequency domain:
\begin{equation} \begin{aligned}
\Delta_{n \Tilde{n}}(\omega) =& \frac{1 + \Bar{n} + i \omega}{1 + \Bar{n} + \Bar{V} \phi^{(1)}(\Bar{V}) + i \omega} \equiv \vcenter{\hbox{
\feynmandiagram [layered layout, horizontal=a to b]
{
  a -- b
};}} \\
\Delta_{V \Tilde{n}}(\omega) =& \frac{-\Bar{V}}{1 + \Bar{n} + \Bar{V} \phi^{(1)}(\Bar{V}) + i \omega} \equiv
\vcenter{\hbox{\feynmandiagram [layered layout, horizontal=a to b]
{
  a -- [photon] b
};}} \\
\Delta_{n \Tilde{V}}(\omega) =& \frac{\phi^{(1)}(\Bar{V})}{1 + \Bar{n} + \Bar{V} \phi^{(1)}(\Bar{V}) + i \omega} \equiv
\vcenter{\hbox{\feynmandiagram [layered layout, horizontal=a to b]
{
  a -- [gluon] b
};}} \\
\Delta_{V \Tilde{V}}(\omega) =& \frac{1}{1 + \Bar{n} + \Bar{V} \phi^{(1)}(\Bar{V}) + i \omega} \equiv
\vcenter{\hbox{\feynmandiagram [layered layout, horizontal=a to b]
{
  a -- [scalar] b
};}}
\end{aligned} \end{equation}
where $\phi^{(n)}(V)$ is the $n$th derivative of the intensity and $\bar{V}$ and $\bar{n}$ are the mean-field estimates of the membrane potential and firing rate, respectively (Eq.~\ref{MFT} for a single population).

The vertices in the Feynman diagrams correspond to source and interaction terms in the action (Appendix \ref{appendix:feynmanrules}). The source terms correspond to sources of fluctuations---here, the stochastic spike emission. 
The interaction terms are generated by the nonlinearities, i.e., the spike reset and the nonlinear intensity function. 
The vertices that give rise to 1-loop corrections are: 
\begin{align}
    \vcenter{\hbox{\begin{tikzpicture}
    \begin{feynman}[small]
        \vertex [dot] (a) {};
        \vertex [above left=of a] (f1) {\(\Tilde{n}\)};
        \vertex [below left=of a] (f2) {\(\Tilde{n}\)};
        \diagram* {
        {(f1), (f2)} -- (a),
        };
    \end{feynman}
    \end{tikzpicture}}}, \quad
    \vcenter{\hbox{\begin{tikzpicture}
    \begin{feynman}[small]
        \vertex [empty square] (a) {};
        \vertex [left=of a] (b) {\(\Tilde{V} \)};
        \vertex [above right=of a] (f1) {\(n\)};
        \vertex [below right=of a] (f2) {\(V\)};
        \diagram* {
        (b) -- (a) -- {(f1), (f2)},
        };
    \end{feynman}
    \end{tikzpicture}}}, \quad
    \vcenter{\hbox{\begin{tikzpicture}
    \begin{feynman}[small]
        \vertex [empty dot] (a) {};
        \vertex [left=of a] (b) {\(\Tilde{n}\)};
        \vertex [above right=of a] (f1) {\(V\)};
        \vertex [below right=of a] (f2) {\(V\)};
        \diagram* {
        (b) -- (a) -- {(f1), (f2)},
        };
    \end{feynman}
    \end{tikzpicture}}}
\end{align}
where the first vertex is a source vertex that emits 2 response variables $\Tilde{n}$ and has an amplitude $\phi^{(0)}(\bar{V})$, the second vertex comes from the nonlinear spike reset and has an amplitude $-1$, and the last vertex comes from the nonlinear intensity function and has an amplitude $\phi^{(2)}(\bar{V}) / 2$. 
The second and third vertices are interaction vertices. 
Using the Feynman rules derived by \cite{ocker2023slif}, we can evaluate the loop corrections to $V$ and $\Dot{n}$. 
Both quantities receive corrections from the intensity function as well as the spike reset. 
The perturbative 1-loop corrections to the mean-field theory are:
\begin{align}
    \langle V \rangle = \Bar{V} \quad + \quad
    \vcenter{\hbox{\begin{tikzpicture}
    \begin{feynman}
        \vertex (a);
        \vertex [left=of a] (b);
        \vertex [right=of a] (c);
        \diagram* {
          a -- [scalar] b [empty square] -- [photon, half left] c [dot] c -- [half left] b
        };
    \end{feynman}
    \end{tikzpicture}}} \quad + \quad
    \vcenter{\hbox{\begin{tikzpicture}
    \begin{feynman}
        \vertex (a);
        \vertex [left=of a] (b);
        \vertex [right=of a] (c);
        \diagram* {
          a -- [photon] b [empty dot] -- [photon, half left] c [dot] c -- [photon, half left] b
        };
    \end{feynman}
    \end{tikzpicture}}}
\end{align}

\begin{align} 
    \langle \Dot{n} \rangle = \Bar{n} \quad + \quad
    \vcenter{\hbox{\begin{tikzpicture}
    \begin{feynman}
        \vertex (a);
        \vertex [left=of a] (b);
        \vertex [right=of a] (c);
        \diagram* {
          a -- [gluon] b [empty square] -- [photon, half left] c [dot] c -- [half left] b
        };
    \end{feynman}
    \end{tikzpicture}}} \quad + \quad
    \vcenter{\hbox{\begin{tikzpicture}
    \begin{feynman}
        \vertex (a);
        \vertex [left=of a] (b);
        \vertex [right=of a] (c);
        \diagram* {
          a -- b [empty dot] -- [photon, half left] c [dot] c -- [photon, half left] b
        };
    \end{feynman}
    \end{tikzpicture}}}
\end{align}
which can be evaluated using the residue theorem (see \cite{kordovan2020spiketraincumulants, helias2020statfieldtheory, zinnQFT2002} for an introduction to evaluating frequency integrals):
\begin{align}
    \langle V \rangle &= \Bar{V} - \frac{\Bar{V}^{2} \phi^{(0)} \phi^{(1)}}{2 (1 + \Bar{n} + \Bar{V} \phi^{(1)})^{2}} - \frac{\Bar{V}^{3} \phi^{(0)} \phi^{(2)}}{4 (1 + \Bar{n} + \Bar{V} \phi^{(1)})^{2}} \label{Perturb_V} \\
    \langle \Dot{n} \rangle &= \Bar{n} - \frac{\Bar{V}^{2} \phi^{(0)} \phi^{(1) 2}}{2 (1 + \Bar{n} + \Bar{V} \phi^{(1)})^{2}} + \frac{\Bar{V}^{2} (1 + \Bar{n})\phi^{(0)} \phi^{(2)} }{4 (1 + \Bar{n} + \Bar{V} \phi^{(1)} )^{2}}. \label{Perturb_FR}
\end{align} 
As noted earlier, the fluctuation corrections in \eqref{Perturb_V}, \eqref{Perturb_FR} come from two terms. The first correction comes about due to the spike reset, and always suppresses both firing rate and membrane potential. 
The second correction is due to the nonlinear intensity function. 
It can either suppress or promote the firing rate/membrane potential depending on the concavity of $\phi$. 
Furthermore, the second correction appears with opposite sign in the two equations. 
This is because the linear response of the rate to spike fluctuations is is positive, while the linear response of the voltage to spike fluctuations is negative due to the spike reset. 
That term could suppress firing rates while paradoxically increasing the mean membrane potential, or vice versa.
This raises the question: under what conditions do fluctuations promote or suppress the mean membrane potential and firing rate of the network? 

\subsubsection{Fluctuation corrections for threshold power-law intensity}
\begin{figure*}
  \begin{center}
  \includegraphics[scale=0.5]{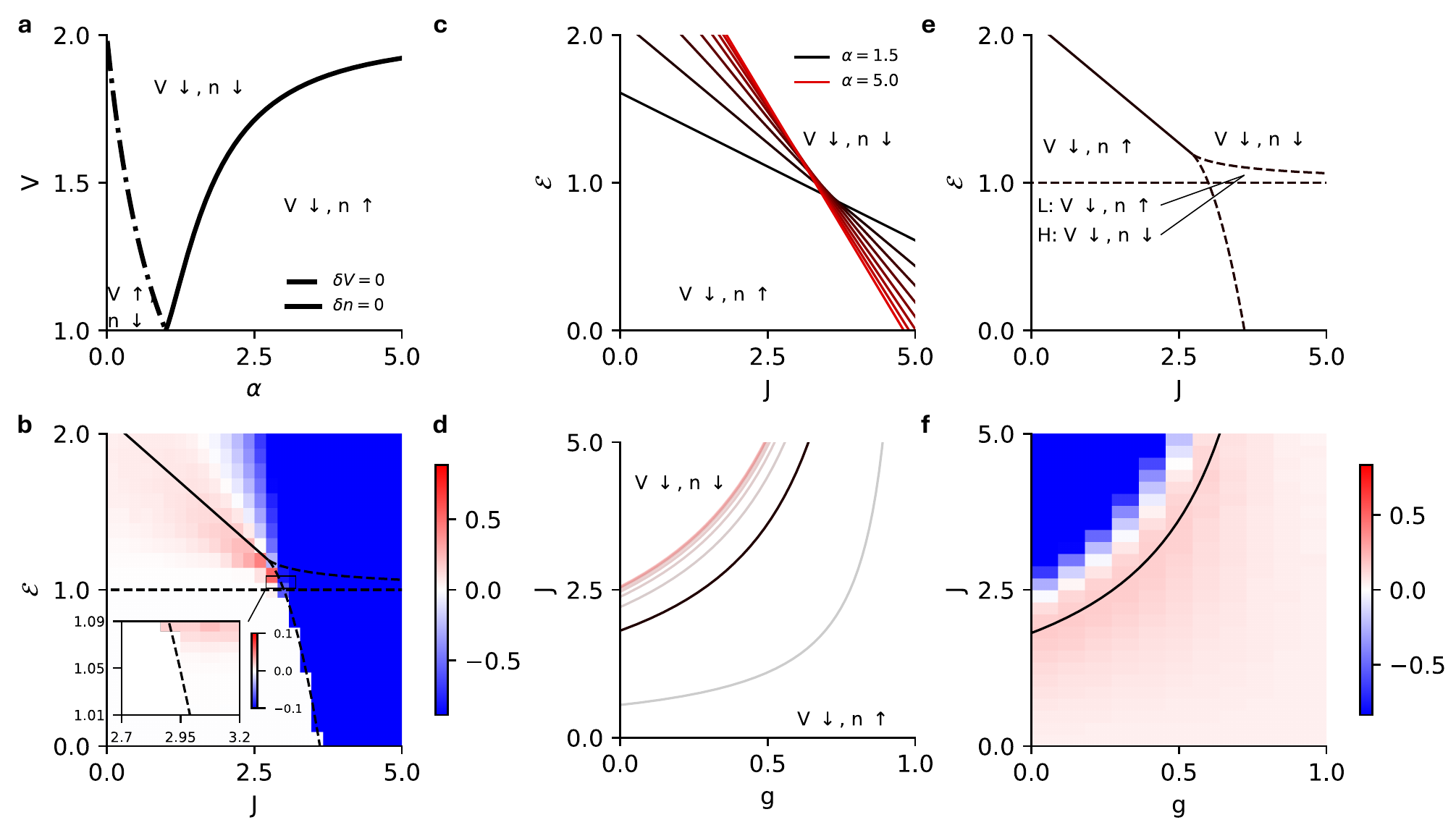}
  \caption{\textbf{a)} Phase diagram in the $V{-}\alpha$ plane showing the regions where fluctuations promote/suppress the mean firing rate and membrane potential. The curves for $\alpha < 1$ and $\alpha > 1$ meet at the same point as $\alpha \rightarrow 1$, where the corrections due to the intensity vanish. \textbf{b} Phase diagram in the $\mathcal{E}{-}J$ plane depicting the lines where the fluctuations vanish for superlinear intensity functions ($\alpha \in [1.5, 5.0]$). Fluctuations promote the firing rate to the left of each line and suppress the membrane potential to the right of each line. The curves reach a limiting line as $\alpha \rightarrow \infty$. \textbf{c)} Combined plot of the phase boundaries (dashed lines) and the vanishing-fluctuation line for $\alpha = 2$. 
  In the bistable region fluctuations promote the firing rate for the low firing rate (L) state and suppress the firing rate for the high firing rate (H) state, hence the vanishing-fluctuation line does not extend into the bistable region. The same is true for each of the lines in \textbf{b}. \textbf{d)} The difference between simulation and mean-field predictions of the firing rate for a homogeneous network with a superlinear intensity ($\alpha=2$). 
  The solid black line represents the theoretical predictions for the points where fluctuations vanish, while the dashed line shows the bistability region. Inset shows the same for the low firing rate regime in the bistable region as in \ref{Figure2} c. \textbf{e)} Same as \textbf{b} but in the $J{-}g$ plane for an EI network. 
  Here, fluctuations promote the firing rate to the right of each line and suppress the firing rate to the left of each line. \textbf{f)} Same as \textbf{d} but in the $J{-}g$ plane for fixed $\mathcal{E}=1.5$.} \label{Figure6}
  \end{center}
\end{figure*}

The threshold power law intensity function, $\phi(V) = \lfloor V-1 \rfloor^\alpha$, can be sublinear ($\alpha<1$) or superlinear ($\alpha>1$).
We find that fluctuations can promote the membrane potential only when the intensity function is sublinear and promote the firing rate only when the intensity function is superlinear. 
The boundary between enhancement versus suppression occurs when the two fluctuation corrections in \eqref{Perturb_V} and \eqref{Perturb_FR} cancel. With $\alpha \neq 1$ there are two mutually exclusive possibilities: the corrections to the mean membrane potential vanish or the corrections to the firing rates vanish. These occur when:
\begin{equation}
    \begin{cases}
        \Bar{V} = \frac{2 \theta}{\alpha + 1} &;~\delta V = 0, \alpha < 1 \\
        \Bar{V} = \Big(\frac{\alpha - 1}{\alpha + 1}\Big)^{1/\alpha} + \theta &;~\delta \Dot{n} = 0, \alpha > 1.
    \end{cases}
\end{equation}
These curves define the surfaces where the one-loop fluctuation corrections vanish (Fig. \ref{Figure6}a). These can be viewed in the $\mathcal{E}{-}J$ plane by substituting the above solutions for $\Bar{V}$ in the mean-field equations of motion \eqref{MFT}:
\begin{equation} \label{EJPlaneFluctuations}
    \begin{cases}
        \mathcal{E} + J \Big(\frac{\alpha + \theta}{\alpha + 1}\Big)^{\alpha} - \Big[ \Big(\frac{\alpha + \theta}{\alpha + 1}\Big)^{\alpha} + 1 \Big] \Big(\frac{2 \theta}{\alpha + 1} \Big)  = 0 &;~\delta V = 0  \\
        \mathcal{E} + J \Big(\frac{\alpha - 1}{\alpha + 1}\Big) - \Big[\Big(\frac{\alpha - 1}{\alpha + 1}\Big)^{1/\alpha} + \theta\Big] \Big(\frac{2\alpha }{\alpha + 1}\Big) = 0 &;~\delta \Dot{n} = 0.
    \end{cases}
\end{equation}  
For the monostable regimes, these are lines in the $\mathcal{E}{-}J$ plane that separate the regions where the membrane potential/firing rate are promoted/suppressed. 
Because our fits in Fig. \ref{Figure1}c find only superlinear nonlinearities, we focus on the $\alpha > 1$ case. 
Increasing $\alpha$ then increases the magnitude of the slope as well as intercept of the lines (Fig. \ref{Figure6}b, c). 
The equations suggest that for a stronger coupling and a larger external input, fluctuations always suppress the firing rate and the membrane potential. 
This is a result of the nonlinear spike reset that prevents the dynamics from exploding despite the unbounded nonlinearity. 
Fig.~\ref{Figure6}d shows the difference between the firing rate of the stochastic spiking network and the mean-field prediction of the firing rate.
\begin{figure}
  \begin{center}
  \includegraphics[width=\columnwidth]{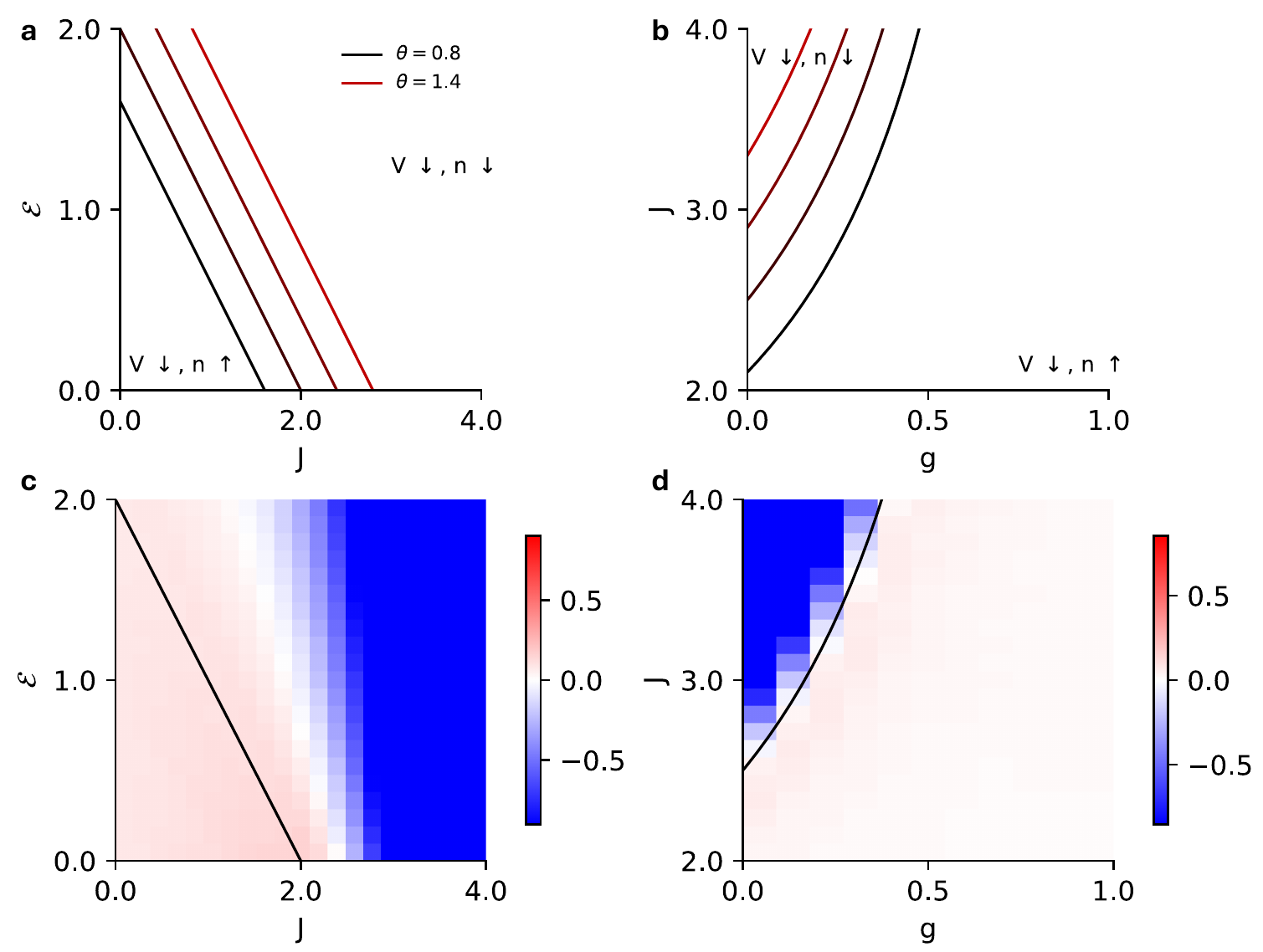}
  \caption{\textbf{a)}  Phase diagram in the $\mathcal{E}{-}J$ plane depicting the lines where the fluctuations vanish for exponential intensity functions ($\theta \in [0.8, 1.4]$). Fluctuations promote the firing rate to the left of each line and suppress the firing rate to the right of each line. \textbf{b)} Same as \textbf{a} but in the $J{-}g$ plane for for an EI network for fixed $\mathcal{E}=-0.5$. Here, fluctuations promote the firing rate to the right of each line and suppress the firing rate to the left of each line. \textbf{c)} The difference between simulation and mean-field predictions of the firing rate for a homogeneous network with exponential intensity ($\theta=1$). The solid black line represents the theoretical predictions for the points where fluctuations vanish. \textbf{d)} Same as \textbf{c} but in the $J{-}g$ plane for fixed $\mathcal{E}=-0.5$.} \label{Figure7}
  \end{center}
\end{figure} 

For the EI network, the same boundaries can be estimated with the replacement $J \rightarrow J(1-g)$ in \eqref{EJPlaneFluctuations}. 
The boundaries for superlinear intensity functions and the comparisons with simulation are shown in Figs.~\ref{Figure6}e,f. 
The equations suggest that--similar to the homogeneous case--fluctuations always suppress the firing rates and membrane potentials for the parameter regimes where the network activity becomes very large.

\subsubsection{Fluctuation corrections for exponential intensity}
The exponential intensity functions receive the same fluctuations corrections as in \eqref{Perturb_V} and \eqref{Perturb_FR}. 
For an exponential, however, $\phi^{(2)}(V) > 0$ for all $V$. 
Hence, there are no parameter regimes where the membrane potential can be promoted by the fluctuations, since both correction terms to $\Bar{V}$ come with a negative sign. 
As before, we assume the intensity function has the form $\phi(V) = e^{V - \theta}$
where $\theta$ is the soft threshold of the intensity function. 
The nullcline where the fluctuations corrections vanish for an exponential intensity function is:
\begin{equation} \label{EJFluctuationsExp}
    \mathcal{E} + J = 2 \theta
\end{equation} 
which again is a straight line in the $\mathcal{E}{-}J$ plane (Fig. \ref{Figure7}a). 
For the EI network, the same boundaries can be estimated with the replacement of $J \rightarrow J(1-g)$ in \eqref{EJFluctuationsExp} (Fig. \ref{Figure7}b). 
Fig.~\ref{Figure7}c, d show the difference between the firing rates of the stochastic spiking network and the mean-field prediction for the homogeneous population ($\mathcal{E}{-}J$ plane) and the EI populations ($J{-}g$ plane) for fixed $\mathcal{E}$, respectively.

\section{Discussion}  \label{Discussion}
We analyzed how single neuron nonlinearities impact overall population activity in networks of stochastic leaky integrate and fire neurons. Here, we specifically focused on the relative roles of the nonlinear spike reset and nonlinear spike intensity functions: the threshold-power law intensity functions that have been fit well to experimentally recorded activity in the visual cortex \cite{miller2002powerlaw, persi2011powerlawvisual, nicholas2008powerlawvisual, hansel2002catvisual}, and the exponential intensity functions that have been extensively used in fitting point process GLMs to neuronal recordings \cite{badel2007exp, pillow2018expGLM, mensi2011expGLMadap, pozzorini2015expGLM, teeter2018expGLM, brinkman2018expGLM, simoncelli2004expGLM, paninski2004MLEexpGLM, pillow2005expGLM, liang2024statistically}. 
We demonstrated that networks with superlinear threshold-power law intensity functions can have a maximum of 2 metastable steady states in the weak coupling regime ($J_{ij} \sim 1/N$). 
Similar to threshold-linear intensity functions \cite{ocker2023slif}, the homogeneous and EI networks can be monostable (active or quiescent), or metastable between a high firing rate and quiescent state. 
However, for superlinear intensities a new metastable region that admits active high- and low-firing rates emerges, similar to that of hard-threshold integrate-and-fire networks \cite{brunel2000dynamics, tartaglia2017updown}, and reminiscent of the up-down transitions seen in numerous cortical areas \cite{kajikawa2022udhippocampus, anderson2000udcatvisual, ilan1999udcatvisual, steriade2001ud, bilal2007udvisual}. 
We find that increasing the exponent of the intensity function continuously expands the region where these active states co-exist. (Fig. \ref{Figure2}). 
This high-low metastability also appears in networks with exponential intensity functions (Fig. \ref{Figure3}), owing to the fact that the firing intensity is never zero{---}the ``subthreshold'' firing probability may be small but remains positive. 
Transitions between the two active states can occur stochastically, or be driven by an external input.

While the mean-field theory provides a decent qualitative description of the network dynamics, it neglects all fluctuations. 
Using a diagrammatic expansion, one can add the effects of nonlinear fluctuations on the network activity. 
In the sLIF model, the corrections to the mean-field come from the nonlinear intensity function and the hard reset of the membrane potential after a spike \cite{ocker2023slif}. 
The effect of the nonlinear intensity depends on its concavity. 
In the monostable regime, for the threshold power law intensity, when the intensity function is sublinear ($\alpha < 1$) fluctuations always suppress the firing rate and can promote/suppress the membrane potential. 
When the intensity function is superlinear ($\alpha > 1$){---}thought to be the more relevant case biologically{---}fluctuations can promote/suppress the firing rate but always suppress the membrane potential.
The two corrections to both variables (firing rate and membrane potential) can vanish simultaneously only when the intensity is threshold linear $\alpha=1$. 
For the exponential intensity, the concavity is always positive, so that the nonlinear intensity always suppresses the membrane potential and always promotes the firing rate. 

Stochastic transitions in multistable firing states have been ubiquitously observed in experimental recordings, and are often associated with behavioral readouts in the tasks \cite{kajikawa2022udhippocampus, holcman2006emergenceUD, mejias2010udnoise, anderson2000udcatvisual, ilan1999udcatvisual, Mazzucato2015MS, mazzucato2019MS, lacamera2019MS, miller2010MStaste, recanatesi2022MS}. 
Understanding how network parameters can shape such stochastic transitions in multistable networks of spiking neurons has been a long standing question in neuroscience \cite{litwinkumar2012MS, schaub2015SSA}. 
The field theoretic formulation provides a powerful method to investigate these transitions. 
Previous work has estimated transition probabilities in networks of binary units (loosely interpreted as active or inactive neurons) \cite{bresloff2010MS, bresslofff2014transitionmarkov, bressloff2010ratemarkov}. 
Adaptations of this formalism can be potentially be applied to these networks of spiking neurons to delineate the underlying mechanisms that drive these stochastic transitions and elucidate how network parameters shape the transition rates between different states. 
We leave this as a direction for future work.

Another avenue for extension of our work is the investigation of networks with ``strong'' synaptic weights that scale as $1/\sqrt{N}$ rather than $1/N$. In strongly coupled networks the variance of the inputs to a neuron is $\mathcal O(1)$, and typically balanced in such a way that the mean input cancels out. 
Firing rate networks with strong synaptic connections have been studied extensively \cite{kadmon2015transitionchaos, sompolinsky1988chaosRNN, schuecker2018optimalRNN, keup2021transchaoticRNN, muscinelli2019singleneuronchaos}, but extending these methods to networks of spiking neurons is non-trivial due to the feedback between the spiking and membrane potential evolution. 
Extending the analysis to a finite-network size and strong synaptic connections could yield valuable insights about the mechanisms that drive stochastic transitions between metastable states.

\begin{acknowledgments}
SP and BAWB thank the National Institute of Mental Health and National Institute for Neurological Disorders and Stroke grant UF-1NS115779-01 and Stony Brook University for financial support for this work.
\end{acknowledgments}

\appendix

\section{Fits to data} \label{appendix:fitting}

We use publicly available cell data from the Allen Institute for Brain Science to fit the nonlinear intensity functions $\phi(V)$ \cite{teeter2018expGLM}. We use data from 6 different cell types from mouse visual cortex; corresponding Cre-line, area, cortical layer, and dendritic type are given in Table~\ref{table0}. 
In the `long square' stimulation protocol, the cells were stimulated at various levels of input current amplitudes to get their steady-state response. We use the data from this stimulation protocol to fit the mean firing rate during the stimulation period as a function of the mean (time-averaged) membrane potential during the inter-spike intervals (excluding the hyperpolarization period), minimizing the squared distance between the data and fit. Within the mean-field approximation we use in this work, a scatterplot of these means is expected to follow $\phi(V)$. The general form of the intensity we consider for the fits are:
\begin{align}
    \begin{cases}
        \phi(V) = \lfloor \gamma (V - \theta) \rfloor_{+}^{\alpha} & \rm{Power ~ Law} \\
        \phi(V) = \exp{[\gamma(V-\theta)]} & \rm{Exponential}
    \end{cases}
\end{align}
where $\gamma$ is the gain of the intensity. The details for the fit are as follows:
\begin{enumerate}
    \item For each level of stimulation, we extracted the mean membrane potential of the cell, excluding the membrane potential values between the time of reaching the threshold (extracted from the spike feature data) and the time of reaching the trough (trough time).
    \item We estimated the firing rate by counting the number of spikes during the stimulation divided by the total stimulation time. 
    \item For the power law intensity, we fixed the threshold at the maximum value of the mean membrane potential where the firing rate was zero. We then fit the gain ($\gamma$) and exponent ($\alpha$).
    \item For the exponential intensity, we treated both gain ($\gamma$) and threshold ($\theta$) as free parameters in our fit.
\end{enumerate}

\begin{table}
\centering
\begin{tabular}{|c|c|c|c|c|}\hline
 \makecell{\textbf{Cell} \\ \textbf{Type \#}} & \textbf{Cre Line} & \textbf{Brain Area} & \textbf{Layer} & \makecell{\textbf{Dendritic} \\ \textbf{Type}} \T\B \\
\hline
 1 & Ctgf-T2A-dgCre & Lateral visual & 6a/b & All \T \\
\hline
 2 & Scnn1a-Tg2-Cre & Primary visual & 5 & Spiny \T \\ 
\hline
 3 & Scnn1a-Tg2-Cre & Primary visual & 4 & Spiny \T \\ 
\hline
 4 & Vip-IRES-Cre & Primary visual & 6a & All \T \\
\hline
 5 & Ctgf-T2A-dgCre & Primary visual & 6b & Spiny \T \\
\hline
 6 & Sim1-Cre-KJ18 & Lateral visual & 5 & Spiny \B \\
\hline
\end{tabular}
\caption{\label{table0} Cre Line, brain area, cortical layer, and dendritic type for each `Cell Type $\#$' label in \ref{Figure1}. Dendritic type `All' includes spiny, aspiny and sparsely spiny types. All cells are from mouse brains. Vip cells (type 4) are inhibitory and mainly target other inhibitory cells \cite{rudy2011three}. All other Cre lines here label various excitatory neurons. }
\end{table}

\section{Steady-states for Threshold Power Law Intensity Function} \label{appendix:sspowerlaw}

Let the right-hand-side of the mean-field equation of motion, assuming a superthreshold mean membrane potential, be
\begin{equation}
    f(\Bar{V}) = -\Bar{V} + \mathcal{E} + (J - \Bar{V}) (\Bar{V} - 1)^{\alpha}
\end{equation}
where we have set the threshold to $\theta = 1$. The values of $\bar{V}$ where $f(\bar{V})$ vanishes correspond to the steady states of the mean-field dynamics.
The solutions cannot generally be obtained in closed form, but we can show that at most two solutions exist that satisfy $\bar{V} \geq 1$.
We do so by investigating the points of inflection of $f(\bar{V})$, looking at the zeros of the second derivative:
\begin{equation}
    f''(\Bar{V}) = -(\Bar{V} - 1)^{\alpha-2} [\bar{V} (-\alpha (\alpha+1)) + \alpha(\alpha-1)J + 2\alpha]
\end{equation}
From this we see that there are only two points of inflection: $\Bar{V}_{{\rm inf},1}=1$ or $\Bar{V}_{{\rm inf},2} = \frac{(\alpha-1)J + 2}{\alpha+1}$. Thus, there is only one point of inflection above threshold for a general $\alpha$. We also see that $\lim_{V\to\infty} f(V) \rightarrow -\infty$. If $\mathcal{E}>1$, the value of the function at $\Bar{V}_{{\rm inf},1}$, $f(\Bar{V}_{{\rm inf},1})$ is positive, which implies the function can have a maximum of 3 roots above threshold. If $\mathcal{E}<1$ then $f(\Bar{V}_{{\rm inf},1})<0$, and the function can have a maximum of 2 roots above threshold. In both cases ones of those roots is an unstable fixed point of the mean-field dynamics, and will not be observed in practice. Accordingly, the mean-field steady-state equation can have a maximum of two stable steady states.

We can also use this to determine the location of the critical point where the two roots meet for a general $\alpha$. The critical point is the point where the first three derivatives of the steady-state equation vanish simultaneously ($f(\Bar{V}) = f'(\Bar{V}) = f''(\Bar{V}) = 0$). This gives
\begin{align}
    J &= 1 + \Big( \frac{\alpha - 1}{\alpha + 1} \Big)^{\frac{1-\alpha}{\alpha}}, \nonumber \\
    \mathcal{E} &= 1 + \Big( \frac{\alpha - 1}{\alpha + 1} \Big)^{\frac{1+\alpha}{\alpha}}.
\end{align}
From this we see that $\lim_{\alpha \rightarrow 1} [J, ~\mathcal{E}] = [2.0, ~1.0]$, and $\lim_{\alpha \rightarrow \infty} [J,~ \mathcal{E}] = [2.0, ~2.0]$. The critical point continuously increases for $\alpha > 1$, thus, expanding the region above threshold where the active-active states can co-exist.

For specific values of $\alpha$ we can obtain exact solutions for the mean-field phase boundaries. In the case $\alpha = 2$ the roots of the fixed point equation
$$\bar{V}^3 - \bar{V}^2(J+2) + \bar{V}(2J+2) -(J + \mathcal E) = 0$$
can be shown to have $1$ or $3$ real-valued roots. The transition from $1$ to $3$ can be determined by investigating the discriminant, a quadratic equation in $\mathcal{E}$:
\begin{align}
    \mathbb{D}_{3} &= -27 \mathcal{E}^{2} + A(J)\mathcal{E}  +B(J) \label{DiscQuad}
\end{align}
where $A(J)=[36(J+2)(J+1) - 4(J+2)^{3} - 54 J]$ and $B(J)=[36J(J+1)(J+2) - 4J(J+2)^{3} + 4(J+2)^{2}(J+1)^{2} - 32(J+1)^{3} - 27J^{2}$. 
The roots of this quadratic equation give us the mean-field boundaries that separate bistable and monostable phases. 

For $\alpha = 3$ the root equation is quartic in $\bar{V}$,
\begin{align}
    \Bar{V}^{4} + b\Bar{V}^{3} - c\Bar{V}^{2} + d \Bar{V} + f &= 0
\end{align}
where the coefficients $\{b,c,d, f\}$ are $b = -(J+3),~c = 3(J+1),~d = -3J,~f = (J + \mathcal{E})$.
The discriminant is cubic in $\mathcal E$:
\begin{equation}
    \mathbb{D}_{4} = -256 \mathcal{E}^3 + \mathcal{E}^{2} A(J) - \mathcal{E} B(J) + C(J) \label{DiscCubic}
\end{equation}
where $A(J) = (768 J - 192bd - 128c^{2} + 144b^2c - 27b^4), B(J) = 768 J^{2} + 2 J [- 192bd - 128c^{2} + 144b^2c - 27b^4] + [144cd^2 - 6b^2d^2 - 80bc^2d + 16c^4 + 18b^3cd - 4b^2c^3], C(J) = 256 J^{3} + J^{2} [- 192bd - 128c^{2} + 144b^2c - 27b^4] + J [144cd^2 - 6b^2d^2 - 80bc^2d + 16c^4 + 18b^3cd - 4b^2c^3]  + [18bcd^3 - 27d^4 - 4c^3d^2 - 4b^3d^3 + b^2c^2d^2]$.
The roots of this cubic equation, along with the condition that $\bar{V}$ must be above threshold and the following inequalities:
\begin{align}
    8c - 3 b^{2} &< 0, \nonumber \\
    64f - 16c^2 + 16b^2c - 16bd - 3b^4 &< 0,
\end{align}
define the boundaries separating the bistable and the monostable phases.

\section{Exact firing rates from renewal theory} \label{appendix:renewalFR} 

After averaging over the synaptic weights $J_{ij}$ and neglecting variance in the weights, which is negligible when $N$ is large, we obtain a population-averaged stochastic differential equation that follows from the action Eq.~\eqref{MGF_DO}:
\begin{align}
    \dot{V}_\mu &= -V_\mu(t) + \mathcal E_\mu - \dot{n}_\mu(t^+)V_\mu(t) + \sum_{\nu = 1}^N J_{\mu \nu} \left\langle \dot{n}_\nu(t)\right\rangle, \label{DOmem}\\
    \dot{n}_\mu(t)dt &\sim {\rm Poiss}\left[\phi(V_\mu(t))dt\right] \label{DOspikes},
\end{align}
where $\langle \dot{n}_\mu(t)\rangle$ is the population-averaged input to each cluster. In the large $N$ limit this concentrates to a constant value that must be determined self-consistently. In this limit the clusters are formally decoupled and interact only through the self-consistent input.

Due to the hard reset of the membrane potential after a spike, each cluster's spike train can be treated as a renewal process. Between spikes the membrane dynamics of each cluster evolves independently according to:
\begin{equation}
    V_\mu(t) = C_\mu (1-e^{-t})
\end{equation}
where $C_\mu = \mathcal{E}_\mu + \sum_{\nu=1}^M J_{\mu \nu} \langle \Dot{n}_\nu \rangle$ is the net input that drives the membrane potential of the cluster $\mu$. 
The instantaneous firing intensity, from the model definition is simply $\phi(V_\mu(t))$. The inter-spike interval density is then the product of the instantaneous firing intensity and the survival probability independently for each cluster:
\begin{equation}
    p_\mu(s) = \phi(V_\mu(s)) \exp{\Bigg( -\int_{0}^{s} \phi(V_\mu(t)) dt \Bigg)}.
\end{equation}
The mean inter-spike interval can then be evaluated as $\langle s_\mu \rangle = \int_{0}^{\infty} ds~s p_\mu(s)$, and the mean firing rate is the inverse of the mean inter-spike interval $\langle \Dot{n}_\mu \rangle = 1 / \langle s_\mu \rangle$. For the nonlinearities we consider here, it is not possible to get a closed form expression for the density. However, we can take a semi-analytic approach to estimate the exact firing rates for the two nonlinearities using renewal theory. In the following we drop the $\mu$ subscripts in the intermediate results for simplicity, as different clusters interact only through the coefficients $C$.

\subsection{Threshold Power Law Intensity}
The inter-spike interval density for the threshold power law intensity function is:
\begin{equation} \label{ISI_TPL}
    p(s) = \begin{cases}
        0 &;~s \leq \ln{\frac{C}{C-\theta}} \\
        \big(C(1-e^{-s}) - \theta \big)^{\alpha}  \\
        \times \exp{\Big[-\int_{0}^{s} \big(C(1-e^{-t}) - \theta \big)^{\alpha} dt \Big]} &;~s > \ln{\frac{C}{C-\theta}}.
    \end{cases}
\end{equation}
While it is not possible to get a closed form expression for this integral, the integral in the exponential of Eqn. \eqref{ISI_TPL} can be evaluated in closed form:
\begin{equation*}
    I_{p}(s|C) \equiv \int_{0}^{s} \big(C(1-e^{-t}) - \theta \big)^{\alpha} dt
\end{equation*}
To get a closed form expression for this integral, we make a change of variables to $y = C (1-e^{-t})- \theta$, after which the integral can be identified as proportional to a representation of the hypergeometric function:
\begin{align}
    I_{p}(s|C) &= \frac{(C(1-e^{-s}) - \theta)^{\alpha+1}}{(C-\theta)(\alpha+1)} \nonumber \\
    &\times \; {}_2F_{1}\Big(1, \alpha+1, \alpha+2, \frac{C(1-e^{-s}) - \theta}{C-\theta}\Big).
\end{align}
The mean firing rate can then be estimated by numerically finding the roots of the following system of equations, restoring the cluster subscripts and recalling that $C_\mu = \mathcal{E}_\mu + \sum_{\nu = 1}^M J_{\mu \nu} \langle \dot{n}_\nu \rangle$:
\begin{equation}
    \langle \Dot{n}_\mu \rangle = \Bigg( \int_{\ln{\frac{C_\mu}{C_\mu-\theta}}}^{\infty} ds~ s \big(C_\mu(1-e^{-s}) - \theta \big)^{\alpha} e^{-I_{p}(s|C_\mu)}\Bigg)^{-1}.
\end{equation}

\subsection{Exponential Intensity}
The inter-spike interval density for the exponential intensity function is:
\begin{equation} \label{ISI_Exp}
    p(s) = \begin{cases}
            0 &;~s \leq 0 \\
            e^{(C (1-e^{-s})- \theta)} \times \\
             \exp{\Big(-\int_{0}^{s} e^{(C (1-e^{-t})- \theta)} dt\Big)} &;~s > 0.
    \end{cases}
\end{equation}
 Again, the integral in the exponential of Eqn. \eqref{ISI_Exp} can be evaluated in closed form by making a change of variables $y = Ce^{-t}$, identifying the result in terms of the ``Exponential Integral'' special function:
\begin{equation}
    I_{e}(s|C)=e^{(C-\theta)} [-{\rm Ei}(-C e^{-s}) + {\rm Ei}(-C)].
\end{equation}
where
${\rm Ei}(x) = -\int_{-x}^{\infty} dt~e^{-t}/t$ is the exponential integral. The mean firing rate can then be estimated by numerically finding the roots of the following system of equations (after restoring the explicit subscipts):
\begin{equation}
    \langle \Dot{n}_\mu \rangle = \Bigg( \int_{0}^{\infty} ds~ s e^{(C_\mu (1-e^{-t})- \theta)} e^{-I_{e}(s|C_\mu)}\Bigg)^{-1}.
\end{equation}

\section{Steady-states for exponential intensity functions} \label{appendix:ssexponential}

The steady-state equation for the exponential intensity function is:
\begin{equation}
    f(\Bar{V}) = -\Bar{V} + \mathcal{E} + (J - \Bar{V}) e^{\Bar{V} - \theta} \label{fExp}
\end{equation}
We can find the extremum of the function by demanding that its derivative vanishes:
\begin{align}
    f'(V) &= -1 - e^{\Bar{V} - \theta} + (J-V) e^{\Bar{V} - \theta} \\
    \Bar{V}_{0,-1} &= (J-1) + W_{0,-1} \left(-e^{\theta + 1 - J}\right) \label{fPrimeExp}
\end{align}
where, $W_{0,-1}$ are the two real branches of the Lambert W function. Thus, the derivative can vanish at either 2 points, or does not vanish at all. Since the argument is the negative of an exponential, the derivative can vanish at exactly 2 points if $-1/e < -e^{\theta + 1 - J} < 0$; otherwise the function is strictly decreasing. Therefore, the necessary condition for the derivative to vanish at 2 points is:
\begin{equation}
    J > \theta + 2
\end{equation}
Since the function can have either 2 extrema or no extrema at all, the mean-field equations can have 3 real roots if the values of the function $f(V)$ at $\Bar{V}_{0,-1}$ have opposite signs, and will have only 1 real root if the values of the function $f(V)$ at $\Bar{V}_{0,-1}$ have the same sign or if $f'(\Bar{V})$ never vanishes ($J < \theta + 2$). Substituting the solution \eqref{fPrimeExp} in \eqref{fExp}:
\begin{equation}
    f(\Bar{V}_{0,-1}) = (\mathcal{E} - J) + \big(1 - W_{0,-1} \big) \big(1 + e^{J-1-\theta+W_{0,-1}} \big)
\end{equation}
where we have suppressed the argument of Lambert W function. Thus, there exist 3 steady-states in the network if $f(\Bar{V}_{0}) > 0$ and $f(\Bar{V}_{-1}) < 0$, or:
\begin{align*}
    \mathcal{E} &> J - (1 - W_{0} ) (1 + e^{J-1-\theta+W_{0}} ) \\
    \mathcal{E} &< J - (1 - W_{-1}) (1 + e^{J-1-\theta+W_{-1}} ).
\end{align*}
The mean-field dynamics are bistable between the two curves above, and monostable otherwise.

\section{Feynman rules for fluctuation corrections} \label{appendix:feynmanrules}

If the fluctuations are weak, we can expand the MGF in \eqref{MGF_DO} in a perturbative series that adds corrections to the mean-field predictions. This can be formally achieved by splitting the action into two parts: i) the free part ($S_{F}$) with terms that are up to bi-linear in the fields $[\Tilde{V}, V, \Tilde{n}, \Dot{n}]$; ii) the interaction part ($S_{I}$) containing the terms with higher powers of the corresponding fields. Splitting the fields in a background mean and fluctuations ($V \rightarrow \Bar{V} + \delta V, \Dot{n} \rightarrow \Bar{n} + \delta \Dot{n}$), we can find the free and interacting part of the action in the MGF for the homogeneous case in \eqref{MGF_DO}:
\begin{align}
    S &= S_{F} + S_{I}, \nonumber \\
    S_{F} &= \Tilde{V} [\partial_{t}{\Bar{V}} + \Bar{V} - \mathcal{E} + J \langle \Dot{n} \rangle - \Bar{n} \Bar{V}] + \Tilde{n} [\Bar{n} - \phi^{(0)}(\Bar{V})], \nonumber \\
    & \Tilde{V} [\partial_{t} + 1 + \Bar{n}] \delta V + \Tilde{V} [\Bar{V}] \delta \Dot{n} + \Tilde{n} \delta \Dot{n} - \Tilde{n} [\phi^{(1)}(\Bar{V})] \delta V, \nonumber \\
    S_{I} &= \Tilde{V} \, \delta \Dot{n} \, \delta V - \sum\limits_{\substack{p=q=0 \\ \setminus \{p=1,q=0\} \\ \setminus \{p=1,q=1\}}}^{\infty} \frac{\Tilde{n}^{p}}{p!} ~ \frac{\phi^{(q)}(\Bar{V})}{q!} (\delta V)^{q}. \label{FIaction}
\end{align}
The MGF can now be expanded around the solution to the mean-field theory in the first line of the free action above:
\begin{align*}
    \frac{d\Bar{V}}{dt} &= -\Bar{V} + \mathcal{E} + J \langle \Dot{n} \rangle - \Bar{n} \Bar{V}, \\
    \Bar{n} &= \phi^{(0)}(\Bar{V}).
\end{align*}
The perturbative expansion uses the fact that the case of a Gaussian action ($S_{F}$) can be solved exactly. The interacting action ($S_{I}$) is then expanded in a functional Taylor series around this Gaussian solution to systematically add the effect of non-Gaussian features of the distribution. An arbitrary moment can be evaluated using:
\begin{align}
    \langle V \rangle &= \Bar{V} + \langle \delta V \rangle, \nonumber \\
    \langle \Dot{n} \rangle &= \Bar{n} + \langle \delta \Dot{n} \rangle, \nonumber \\
    &\Big\langle \prod_{i=1}^{a} \delta \Dot{n}(t_{i}) \prod_{k=1}^{b} \delta V(t_{k}) \Big\rangle \nonumber \\ 
    &= \prod_{i=1}^{a} \prod_{k=1}^{b} \frac{1}{Z[0]} \frac{\delta}{\delta \tilde{h}(t_{i})} \frac{\delta}{\delta \tilde{j}(t_{k})} Z[\Tilde{j}, j, \Tilde{h}, h]
\end{align}
where $Z[0]$ is the MGF evaluated at vanishing source terms, equal to $1$ because of normalization of the probability density. 
At the lowest order in the perturbation, the expectation value of the fluctuations is zero $(\langle \delta V \rangle = \langle \delta \Dot{n} \rangle = 0)$, so that $(\langle V \rangle = \Bar{V}, \langle \Dot{n} \rangle = \Bar{n})$. This is the mean-field solution. The loop expansion gives these fluctuations a non-zero expectation which provides fluctuation corrections to the mean-field solution. The MGF can now be expanded around the free part of the action to calculate expectations beyond the lowest order:
\begin{align}
    Z[\Tilde{j}, j, \Tilde{h}, h] &= \int \mathcal{D}\tilde{V} \mathcal{D} \delta V \mathcal{D} \tilde{n} \mathcal{D} \delta \Dot{n}~e^{-S_{F} - S_{I} + \Tilde{j} \cdot V + j \cdot \Tilde{V} + \Tilde{h} \cdot \Dot{n} + h \cdot \Tilde{n}}, \nonumber \\
    Z[\Tilde{j}, j, \Tilde{h}, h] &= \int \mathcal{D}\tilde{V} \mathcal{D} \delta V \mathcal{D} \tilde{n} \mathcal{D} \delta \Dot{n}~e^{-S_{F}} \Bigg[1 \nonumber \\
    & + \Big(-S_{I} + \Tilde{j} \cdot V + j \cdot \Tilde{V} + \Tilde{h} \cdot \Dot{n} + h \cdot \Tilde{n} \Big) \nonumber \\
    & + \frac{1}{2!} \Big(-S_{I} + \Tilde{j} \cdot V + j \cdot \Tilde{V} + \Tilde{h} \cdot \Dot{n} + h \cdot \Tilde{n} \Big)^{2} \nonumber \\
    & + \frac{1}{3!} \Big(-S_{I} + \Tilde{j} \cdot V + j \cdot \Tilde{V} + \Tilde{h} \cdot \Dot{n} + h \cdot \Tilde{n} \Big)^{3} + \dots \Bigg]
\end{align}
This expansion can be organized diagrammatically using Feynman rules to estimate various moments of a given distribution in terms of the propagators evaluated at the mean-field solution. Using Wick's theorem, any expectation can be broken into a sum over product of propagators $\langle x(t) \Tilde{x}(t')\rangle \rightarrow [\Delta_{n \Tilde{n}}, \Delta_{V \Tilde{n}}, \Delta_{n \Tilde{V}}, \Delta_{V \Tilde{V}}]$ (where $x = V$ or $\dot{n}$ and $\tilde{x} = \tilde{V}$ or $\tilde{n}$). The Ito condition $\langle x(t) \Tilde{x}(t)\rangle = 0$ ensures that the propagators are purely causal in the time domain. Thus, the only terms that survive in the above expansion are the terms that have an even number of response $[\Tilde{V}, \Tilde{n}]$ and field $[V, \Dot{n}]$ variables, and where each variable in these pairs has a different time argument (comes from different term in the expansion). 
\begin{table}
\centering
\begin{tabular}{|c|c|c|c|}\hline
\textbf{Vertex} & \textbf{Factor} & \textbf{In-degree $(\delta V, \delta \Dot{n})$} & \textbf{Out-degree $(\Tilde{V}, \Tilde{n})$} \T\B \\
\hline
\begin{tikzpicture}
    \begin{feynman}
        \vertex [dot] (a) {};
        \diagram* {(a)};
    \end{feynman}
\end{tikzpicture} & $\phi^{(0)}(\Bar{V})$ & $(0, 0)$ & $(0, \geq 2)$ \T \\
\hline
\begin{tikzpicture}
    \begin{feynman}
        \vertex [empty square] (a) {};
        \diagram* {(a)};
    \end{feynman}
\end{tikzpicture} & -1 & $(1, 1)$ & $(1, 0)$ \T \\ 
\hline
\begin{tikzpicture}
    \begin{feynman}
        \vertex [empty dot] (a) {};
        \diagram* {(a)};
    \end{feynman}
\end{tikzpicture} & $\frac{\phi^{(q)}(\Bar{V})}{q!}$ & $(\geq 1, 0)$ & $(0, \geq q)$ \B \\ 
\hline
\end{tabular}
\caption{\label{table1} The vertices in Feynman diagrams correspond to the terms in the interacting part of the action $(S_{I})$. Each vertex can have incoming field variables (in-degree) or outgoing response variables (out-degree).}
\end{table}
\begin{table}
\centering
\begin{tabular}{|c|c|c|}\hline
\textbf{Propagator} & \textbf{Edge} & \textbf{Factor $(t,s)$} \T\B \\
\hline
$\Delta_{n \Tilde{n}}(t,s)$ & \feynmandiagram [layered layout, horizontal=a to b]
{
  a -- b
}; & \makecell{$\delta(t-s)$ \\ $- \Bar{V} \phi^{(1)} e^{-(1 + \Bar{n} + \Bar{V} \phi^{(1)})(t-s)} \Theta(t-s) $} \T \\
\hline
$\Delta_{n \Tilde{V}}(t,s)$ & \feynmandiagram [layered layout, horizontal=a to b]
{
  a -- [gluon] b
}; & $\phi^{(1)} e^{-(1 + \Bar{n} + \Bar{V} \phi^{(1)})(t-s)} \Theta(t-s) $ \T\B \\ 
\hline
$\Delta_{V \Tilde{V}}(t,s)$ & \feynmandiagram [layered layout, horizontal=a to b]
{
  a -- [scalar] b
}; & $e^{-(1 + \Bar{n} + \Bar{V} \phi^{(1)})(t-s)} \Theta(t-s) $ \T \\ 
\hline
$\Delta_{V \Tilde{n}}(t,s)$ & \feynmandiagram [layered layout, horizontal=a to b]
{
  a -- [photon] b
}; & $-\Bar{V}e^{-(1 + \Bar{n} + \Bar{V} \phi^{(1)})(t-s)} \Theta(t-s) $ \B \\ 
\hline
\end{tabular}
\caption{\label{table2} The edges in the Feynman diagrams correspond to the propagators evaluated from the bilinear part of the free action $(S_{F})$. Each propagator measures the linear response of a field variable to perturbations in the other variable, and are causal due to the Ito condition. Here $\Theta(t-s)$ is the Heaviside step function.}
\end{table}
The diagrams are constructed out of vertices and edges. The vertices (Table \ref{table1}) are the terms in the interacting part of the action $(S_{I})$, and the edges (Table \ref{table2}) are constructed out of the propagators evaluated at the solution to the mean-field theory. The diagrammatic rules to evaluate an arbitrary expectation $\langle \delta \dot{n}(t_1) \dots \delta \dot{n}(t_a) \delta V(t_{a+1}) \dots \delta V(t_{a+b})\rangle$ are as follows:
\begin{enumerate}
    \item Place $a+b$ external vertices corresponding to the expectation value that needs to be evaluated.
    \item Use the vertices and edges in Tables \ref{table1}, \ref{table2} to construct all possible connected graphs such that the external vertices have only one incoming propagator. 
    \item The analytic expression for the diagram can be obtained by multiplying all the factors of edges and vertices together. In the frequency domain, each edge is assigned its own frequency variable and the expression can be evaluated by integrating over all internal frequencies. In the time domain, each vertex carries its own time argument, and the propagators come with the time arguments of the vertices they connect. The expression can be evaluated by again integrating over all internal time points.
    \item The resulting expectation is the sum of all such diagrams that can be constructed in (2).
\end{enumerate}
For the action in \eqref{FIaction}, the vertices and the resulting 1-loop diagrams are (diagrams generated with \cite{tikzfeynman}):
\begin{align}
    \langle \delta V \rangle = \quad
    \vcenter{\hbox{\begin{tikzpicture}
    \begin{feynman}
        \vertex (a);
        \vertex [left=of a] (b);
        \vertex [right=of a] (c);
        \diagram* {
          a -- [scalar] b [empty square] -- [photon, half left] c [dot] c -- [half left] b
        };
    \end{feynman}
    \end{tikzpicture}}} \quad + \quad
    \vcenter{\hbox{\begin{tikzpicture}
    \begin{feynman}
        \vertex (a);
        \vertex [left=of a] (b);
        \vertex [right=of a] (c);
        \diagram* {
          a -- [photon] b [empty dot] -- [photon, half left] c [dot] c -- [photon, half left] b
        };
    \end{feynman}
    \end{tikzpicture}}}
\end{align}

\begin{align} 
    \langle \delta \Dot{n} \rangle = \quad
    \vcenter{\hbox{\begin{tikzpicture}
    \begin{feynman}
        \vertex (a);
        \vertex [left=of a] (b);
        \vertex [right=of a] (c);
        \diagram* {
          a -- [gluon] b [empty square] -- [photon, half left] c [dot] c -- [half left] b
        };
    \end{feynman}
    \end{tikzpicture}}} \quad + \quad
    \vcenter{\hbox{\begin{tikzpicture}
    \begin{feynman}
        \vertex (a);
        \vertex [left=of a] (b);
        \vertex [right=of a] (c);
        \diagram* {
          a -- b [empty dot] -- [photon, half left] c [dot] c -- [photon, half left] b
        };
    \end{feynman}
    \end{tikzpicture}}}
\end{align}
The time arguments corresponding to each vertex are: (\begin{tikzpicture}
    \begin{feynman}
        \vertex [dot] (a) {};
        \diagram* {(a)};
    \end{feynman}
\end{tikzpicture} - $t_{1}$), (\begin{tikzpicture}
    \begin{feynman}
        \vertex [empty square] (a) {};
        \diagram* {(a)};
    \end{feynman}
\end{tikzpicture}, \begin{tikzpicture}
    \begin{feynman}
        \vertex [empty dot] (a) {};
        \diagram* {(a)};
    \end{feynman}
\end{tikzpicture} - $t_{2}$). The above diagrams in terms of the propagators and edges are:
\begin{align}
    \langle \delta V \rangle &= - \phi^{(0)} \int_{-\infty}^{t} dt_{1} dt_{2} \Delta_{V \Tilde{V}}(t, t_{1}) \Delta_{V \Tilde{n}}(t_{1}, t_{2}) \Delta_{n \Tilde{n}}(t_{1}, t_{2}) \nonumber \\
    & + \frac{\phi^{(0)} \phi^{(2)}}{2}\int_{-\infty}^{t} dt_{1} dt_{2} \Delta_{V \Tilde{n}}(t, t_{1}) \Delta_{V \Tilde{n}}(t_{1}, t_{2}) \Delta_{V \Tilde{n}}(t_{1}, t_{2}), \\
    \langle \delta \Dot{n} \rangle &= - \phi^{(0)} \int_{-\infty}^{t} dt_{1} dt_{2} \Delta_{n\Tilde{V}}(t, t_{1}) \Delta_{V \Tilde{n}}(t_{1}, t_{2}) \Delta_{n \Tilde{n}}(t_{1}, t_{2}) \nonumber \\
    & + \frac{\phi^{(0)} \phi^{(2)}}{2}\int_{-\infty}^{t} dt_{1} dt_{2} \Delta_{n \Tilde{n}}(t, t_{1}) \Delta_{V \Tilde{n}}(t_{1}, t_{2}) \Delta_{V \Tilde{n}}(t_{1}, t_{2}).
\end{align}
These expression can be evaluated either in the time domain or can be transformed into the frequency domain and evaluated using the residue theorem. For a detailed introduction to expansions of the moment generating functions see \cite{zinnQFT2002, helias2020statfieldtheory}, path integrals specifically applied to stochastic differential equations see \cite{buiceandchow2015pathintegralsdes}, calculations involving loop integrals in the frequency domain see \cite{kordovan2020spiketraincumulants}, perturbative expansion for this specific model and its connections to the effective action see \cite{ocker2023slif}.


%

\end{document}